\documentclass[revtex4]{emulateapj}

\usepackage{amssymb}
\usepackage{amsmath}
\usepackage[]{graphicx}
\usepackage{enumerate}
\usepackage{subeqnarray}
\usepackage{cases}
\usepackage{mathrsfs,amssymb}
\usepackage{xcolor}
\definecolor{vcol}{rgb}{0.1,0.1,0.5}
\definecolor{vcol2}{rgb}{0.9,0.1,0.2}

\newcommand{\tf}{\widetilde\varphi}
\newcommand{\tm}{\widetilde m}

\tightenlines

\begin{document}

\title{Ionization and dust charging in protoplanetary disks}

\author{A.V. Ivlev$^1$, V.V. Akimkin$^2$, P. Caselli$^1$}
\email[e-mail:~]{ivlev@mpe.mpg.de} \affiliation{$^1$Max-Planck-Institut f\"ur Extraterrestrische Physik, Giessenbachstr. 1,
85748 Garching, Germany } \affiliation{$^2$Institute of Astronomy of the Russian Academy of Sciences, Pyatnitskaya St. 48,
119017 Moscow, Russia}

\begin{abstract}
Ionization-recombination balance in dense interstellar and circumstellar environments is a key factor for a variety of
important physical processes, such as chemical reactions, dust charging and coagulation, coupling of the gas with magnetic
field and development of instabilities in protoplanetary disks. We determine a critical gas density above which the
recombination of electrons and ions on the grain surface dominates over the gas-phase recombination. For this regime, we
present a self-consistent analytical model which allows us to exactly calculate abundances of charged species in dusty gas,
without making assumptions on the grain charge distribution. To demonstrate the importance of the proposed approach, we
check whether the conventional approximation of low grain charges is valid for typical protoplanetary disks, and discuss the
implications for dust coagulation and development of the ``dead zone'' in the disk. The presented model is applicable for
arbitrary grain-size distributions and, for given dust properties and conditions of the disk, has only one free parameter
-- the effective mass of the ions, shown to have a low effect on results. The model can be easily included
in numerical simulations following the dust evolution in dense molecular clouds and protoplanetary disks.
\end{abstract}


\maketitle

\section{Introduction}
\label{intro}

An accurate calculation of ionization-recombination balance in dense protoplanetary conditions is essential for
understanding various fundamental problems, such as coupling of the gas with magnetic field \citep{Li2014}, accretion
processes \citep{Turner2014}, chemistry \citep{Semenov2004,Larsson2012} and dust evolution~\citep{Okuzumi2011b,Akimkin2015}.
Both the ionization and recombination processes can arise from several sources. While the treatment of ionization, despite
the variety of ionization sources, could be reduced to a single (total) ionization rate, the description of recombination is
less straightforward. At sufficiently high densities, the dominant sink of free electrons and ions are dust grains, and the
recombination rate non-trivially depends on properties of the grains. Furthermore, collection of electrons and ions leads to
non-zero grain charges, which effectively changes the grain-grain \citep{Okuzumi2009} and ion-grain \citep{Weingartner1999}
interactions as well as the grain dynamics.

Depletion of electrons caused by the presence of dust grains significantly reduces the degree of ionization in dense
interstellar conditions \citep{Umebayashi1983,Umebayashi1990,Nishi1991}: In comparison with dust-free gas, the
electron-to-ion ratio may drop by as much as a square root of the effective ion-to-electron mass ratio (which is a factor of
74 for a plasma with dominant H$_3^+$ ions, or 231 for N$_2$H$^+$/HCO$^+$ ions). As the ionization controls the coupling of
the gas to the magnetic field, and hence the development of the magnetorotational instability \citep[MRI,
e.g.,][]{Velikhov1959,Balbus1991}, dust is the essential ingredient for any MRI model. It has been shown that the grain size
critically affects the size of a disk's ``dead zone'' \citep{Sano2000,Salmeron2008,Bai2011a,Bai2011b,Dudorov2014}.
Nevertheless, analysis of MRI has been usually carried out assuming that properties of dust are fixed.

In dense protoplanetary environments, the coagulation of sub-$\mu$m interstellar dust particles becomes an important
process. The planet formation in protoplanetary disks requires the dust to form larger and larger aggregates, until
gravitational forces become dominant \citep[e.g.,][and references therein]{Testi2014}. There is a clear evidence of grain
growth to millimeter and centimeter sizes within protoplanetary disks \citep[e.g.,][]{Perez2015,vanderMarel2015}. However,
significant difficulties are found during this coagulation process, such as bouncing barriers \citep[e.g.,][]{Zsom2010} and
particle fragmentation \citep[][]{Birnstiel2012} after initial grain compaction and growth. Many theoretical and laboratory
studies have greatly advanced our understanding of grain growth and planetesimal formation in recent years \citep[][and
references therein]{Dominik2007,Johansen2014}, with particular attention dedicated to dust traps, now detected with ALMA
toward protoplanetary disks \citep[][]{vanderMarel2013,vanderMarel2015,Pinilla2015,Flock2015,Zhang2016,Ruge2016}. In dust
traps, particles are expected to grow more easily due to the locally enhanced dust-to-gas mass ratio
\citep[][]{Booth2016,Surville2016}, although the details of this coagulation process are far from being understood,
considering the largely unknown dust properties.

As has been already pointed out, the ionization does not only determine dynamical and chemical processes occurring in
protoplanetary disks, but also leads to the dust charging and thus affects the coagulation. Collection of electrons and ions
results in (on average) negative grain charges due to higher electron velocities. Recently, it has been shown that the
coagulation of larger aggregates in protoplanetary disks can be inhibited due to growing Coulomb repulsion between them~--
the resulting electrostatic potential barrier is roughly proportional to the aggregate size
\citep[][]{Okuzumi2009,Okuzumi2011a,Okuzumi2011b}. Along with the plasma charging, other charging mechanisms can operate in
protoplanetary disks. In \cite{Akimkin2015} the photoelectric emission from grains, induced by stellar radiation and leading
to their positive charging, was considered as a mechanism to overcome the electrostatic barrier in upper disk regions. A
similar mechanism~-- photoelectric charging due to H$_2$ fluorescence induced by cosmic rays (CRs) -- operates in much
deeper regions at the disk periphery \citep{Ivlev2015b}. However, both mechanisms become negligible in dense regions of the
disk. We notice that the (still poorly investigated) effect of charging on the dust evolution has recently received
increased interest \citep{Carballido2016}.

It is noteworthy to mention that the coagulation in protoplanetary disks is accompanied by the formation of porous
aggregates characterized by an open, fluffy structure \citep[e.g.,][]{Dominik2007,Okuzumi2009b}. The
porosity has been pointed out to have a strong impact on the ionization in protoplanetary disks
\citep[][]{Okuzumi2009,Dzyurkevich2013,Mori2016}. However, while the growth of {\it uncharged} aggregates is well studied,
an accurate description of their charging as well as of the charging feedback on their further growth poses a serious
problem. One of the fundamental difficulties is that, unlike compact spherical grains \citep[whose charging is described
using the Orbital Motion Limited (OML) approximation, e.g.,][]{Whipple1981,Fortov2005}, no accurate approximation is known
for the electron and ion collection by irregular fluffy aggregates. Given these difficulties, here we leave completely aside
in-depth discussion of the porosity effects.

In this paper, we present an analytical model which becomes exact in sufficiently dense astrophysical environments and
allows us to self-consistently calculate densities of the charged species, in particular -- to obtain the dust charges for
arbitrary grain-size distributions. Unlike other known approaches
\citep[][]{Ilgner2006,Okuzumi2009,Fujii2011,Dzyurkevich2013,Mori2016}, our model does not make assumptions on the form of
the charge distribution, and yields closed analytical expressions for important limiting cases. The latter enables
convenient analysis of results in a general form, in terms of a few dimensionless numbers.

The presented model has only one free parameter (the effective mass of the ions, which we show to have a low effect on
results), and can be easily included in numerical simulations following the dust evolution in dense molecular clouds and
protoplanetary disks. We employ the model to verify whether the broadly used approximation of low grain charges is valid for
typical protoplanetary disks. Furthermore, we identify a ``dust-dust'' plasma regime, where the grain charge distribution
becomes quasi-symmetric with respect to uncharged state. This leads to removal of the repulsive electrostatic barrier and
opens a ``coagulation window'' for large aggregates, operating in the inner dense region of protoplanetary disks. Also, we
discuss the importance of self-consistent analysis of the ionization and the grain evolution, as there processes are
mutually coupled via several mechanisms operating in the disks.

The paper is organized as follows. In Section~\ref{dense} we consider the overall ionization-recombination balance and
introduce a recombination threshold -- the gas density above which the electron-ion recombination is dominated by the
processes on the dust surface. In Section~\ref{charge} we present the grain charge distribution determined by collection of
electrons and ions, and point out limiting cases of ``big'' and ``small'' grains. In Section~\ref{densities} we derive the
governing equations for the dust-phase recombination regime, complemented with the grain charge distribution, which allow us
to calculate densities of the charged species in a general form; to reveal generic properties of the solution, we consider
``monodisperse'' dust (grains of the same size) and investigate separately the big- and small-grain limits. The effect of
the grain-size distribution is studied in Section~\ref{size_effect}. We discuss implications of the proposed model for
protoplanetary disks in Section~\ref{implications}, and summarize the results in Section~\ref{conclusion}.

\section{Dust-phase recombination regime}
\label{dense}

The ionization-recombination balance for electrons is generally governed by the following equation:
\begin{equation}\label{ion-rec0}
\zeta n_{\rm g}=R_{\rm g}+R_{\rm d},
\end{equation}
where $\zeta$ is the total ionization rate\footnote{The magnitude of $\zeta$ is determined by a combination of different
ionization mechanisms (due to CRs, X-rays, UV, and radionuclides) whose relative importance varies across the disk
\citep[e.g.,][]{Armitage2015}.} of gas with the number density $n_{\rm g}$. The recombination is represented by the two
terms. The first one, $R_{\rm g}= \sum_k\beta^{(k)}_{\rm g}n_{\rm e}n^{(k)}_{\rm i}$, describes the gas-phase recombination
of electrons (with the density $n_{\rm e}$) and ions, where $\beta_{\rm g}^{(k)}$ is the rate of recombination for the $k$th
ion species (with the density $n_{\rm i}^{(k)}$, the summation is over all ion species). The second recombination term,
$R_{\rm d}=\beta_{\rm d}n_{\rm e}n_{\rm d}$, describes the electron collection on dust (with the density $n_{\rm d}$; in
equilibrium, the electron collection is equal to the collection of all ion species). The dust-phase recombination rate
$\beta_{\rm d}=2\sqrt{2\pi}a^2v_{\rm e} e^{-\Psi}$ (where $v_{\rm e}=\sqrt{k_{\rm B}T/m_{\rm e}}$ is the thermal velocity
scale for electrons) is determined by the normalized potential $\Psi=|\langle Z\rangle|e^2/ak_{\rm B}T$ of a grain of radius
$a$ \citep[][]{Fortov2005}. It is assumed that dust comprises a small mass fraction of gas $f_{\rm d}$ (typically, of the
order of $10^{-2}$ in the interstellar medium), i.e., the dust density is proportional to the gas density:
\begin{equation}\label{fraction}
m_{\rm d}n_{\rm d}=f_{\rm d}m_{\rm g}n_{\rm g},
\end{equation}
where $m_{\rm d}$ is the mass of a grain and $m_{\rm g}\simeq2.3m_{\rm p}$ is the mean mass of a gas particle (expressed in
units of the proton mass $m_{\rm p}$). We note that generation of local dust traps in the disk
\citep{vanderMarel2013,Flock2015,Ruge2016} as well as dust settling to the midplane \citep[][]{Zsom2011} may substantially
increase the value of $f_{\rm d}$.

To evaluate the relative contribution of the two recombination terms, one can set $n_{\rm e}= \sum_kn_{\rm i}^{(k)}$ (this
is verified in Section~\ref{tf<<1}). We substitute Equation~(\ref{fraction}) in Equations~(\ref{ion-rec0}) and obtain a
quadratic equation for $n_{\rm e}$. The solution suggests that for sufficiently low $n_{\rm g}$ the gas-phase recombination
is the dominant process (i.e., $R_{\rm d}$ is negligible), and $n_{\rm e}$ varies with the gas density as $\propto
\sqrt{\zeta n_{\rm g}}$.\footnote{When $R_{\rm g}$ includes both the dissociative recombination and the radiative
recombination with heavy metal ions, the scaling for $n_{\rm e}$ may change between $\propto(\zeta n_{\rm g})^{1/3}$ and
$\propto\zeta n_{\rm g}$, depending on the density of metals and the magnitude of $\zeta$. Such a behavior is typical for
different regions of molecular clouds \citep[e.g.,][]{Oppenheimer1974}.} At higher $n_{\rm g}$ the situation is reversed:
when the gas density exceeds a {\it dust-phase recombination threshold},\footnote{The recombination threshold is defined as
the gas density at which $R_{\rm g}(n_{\rm g})=R_{\rm d}(n_{\rm g})$.}
\begin{equation}\label{th1}
n_{\rm g}\gtrsim n_{\rm g}^{\rm rec}=\frac{\zeta\beta_{\rm g}(m_{\rm d}/m_{\rm g})^2}{2\pi f_{\rm d}^2 v_{\rm e}^2 a^4}e^{2\Psi},
\end{equation}
(where $\beta_{\rm g}$ is the characteristic gas-phase recombination rate), the electron/ion density is primarily determined
by the recombination on grains. By substituting typical values $\zeta\sim10^{-17}$~s$^{-1}$ and $T\sim10^2$~K, and taking
into account that $\beta_{\rm g}\lesssim 10^{-7}$~cm$^3$s$^{-1}$ \citep[e.g.,][]{Okuzumi2009}, for micron-size grains we
obtain that the gas-phase recombination in a heavy-ion HCO$^+$/N$_2$H$^+$ plasma ($\Psi=3.86$) is negligible for $n_{\rm
g}\gg n_{\rm g}^{\rm rec}\sim 10^9$~cm$^{-3}$; for a H$_3^+$ plasma ($\Psi=2.94$) the recombination threshold is an order of
magnitude lower. The characteristic recombination rate $\beta_{\rm g}$ depends on details of the gas-phase reactions
\citep[][]{Oppenheimer1974}, and therefore generally the value of $n_{\rm g}^{\rm rec}$ is known only approximately.

Thus, for $n_{\rm g}\gg n_{\rm g}^{\rm rec}$ -- below we call this the {\it dust-phase recombination regime} -- the
ionization degree is (practically) not affected by a variety of reactions occurring in the gas phase. The emergence of this
regime reflects the growing importance of dust in the global ionization-recombination balance. When the grain-size
polydispersity is taken into account, then (depending on the particular shape of the size distribution) the value of $n_{\rm
g}^{\rm rec}$ can be decreased significantly (see Section~\ref{size_effect}).

\section{Grain charge distribution}
\label{charge}

A stationary discrete charge distribution $N(Z,a)\equiv N_{Z}$ is obtained from the detailed equilibrium of the charging
master equation \citep[][]{Draine1987,Draine2011Book}, as presented in Appendix~\ref{charge_dist}. The charge distribution,
derived for dust grains of a given size $a$ and normalized to the total differential dust density at that size,
\begin{equation}\label{norm}
\sum_Z N_{Z}=dn_{\rm d}(a)/da,
\end{equation}
depends on two dimensionless numbers.

The first number is the effective ion-to-electron mass ratio $\tm$, determined by the partial contributions of all ions. It
is defined as \citep[][]{Draine1987,Ivlev2015b}
\begin{equation}\label{parameter1}
\frac{\tm}{1836}=\left(\sum_k\frac1{\sqrt{A_{k}}}\frac{n_{\rm i}^{\left(k\right)}}{n_{\rm e}}\right)^{-2}\equiv A
\left(\frac{n_{\rm e}}{n_{\rm i}}\right)^2,
\end{equation}
where $n_{\rm i}\equiv\sum_kn_{\rm i}^{(k)}$ is the total ion density and $A_{k}$ is the atomic mass number (in amu) of the
$k$th ion species; the identity in Equation~(\ref{parameter1}) introduces the effective atomic mass of ions $A$. The value
of $\tm$ has well-defined upper and lower bounds: When dust does not noticeably affect the overall charge neutrality, we
have $n_{\rm e}\simeq n_{\rm i}$ and therefore $\tm\simeq1836A~(\ggg1)$; when the contribution of the negatively charged
dust is important, electrons become depleted and one can show (see Section~\ref{densities}) that $\tm$ asymptotically tends
to unity (i.e., $n_{\rm e}/n_{\rm i}$ tends to a small constant of $1/\sqrt{1836A}$), in order to satisfy the charge
neutrality.

The second number is the grain potential energy of the unit charge, $e^2/a$, normalized by $k_{\rm B}T$
\citep[][]{Ivlev2015b},
\begin{equation}\label{parameter2}
\tf=\frac{e^2}{ak_{\rm B}T}=\frac{1.67}{(a/0.1~\mu{\rm m})(T/100~{\rm K})},
\end{equation}
\citep[equivalently, $\tf$ is the inverse reduced temperature,][]{Draine1987}. The value of $\tf$ can, in principle, be
arbitrary small or large. The case $\tf\ll1$ corresponds to situations where grains grow beyond several microns or/and the
ambient gas temperature exceeds a few hundred Kelvin -- such conditions are at best matched in the inner midplane region of
the disk \citep[][]{Armitage2007}. The opposite limit, corresponding to small grains with $a\lesssim0.1~\mu$m or/and low
temperatures of $\lesssim30$~K, primarily represents the initial coagulation stage or a stage where the particle
fragmentation barrier is reached after initial grain growth; this may also represent the outer (colder) disk regions, where
the gas density is nevertheless high enough to satisfy condition (\ref{th1}).

>From Equations~(\ref{distr-}) and (\ref{distr}) we obtain the charge distribution for two limiting cases: Irrespective of
the magnitude of $\tm$, for $\tf\ll1$ the distribution tends to a Gaussian form \citep[][]{Draine1987},
\begin{equation}\label{Gauss}
\tf\ll1:\quad N_{Z}\propto e^{-(Z-\langle Z\rangle)^2/2\sigma_{Z}^2},
\end{equation}
the average charge $\langle Z\rangle=-\tf^{-1}\Psi$ and the charge variance $\sigma_{Z}^2=\tf^{-1}(1+\Psi)/(2+\Psi)$ are
determined by the normalized potential $\Psi$, which is the solution of the charging equation,
\begin{equation}\label{Psi}
(1+\Psi)e^{\Psi}= \sqrt{\tm}.
\end{equation}
For $\tf\gg1$ the distribution is essentially discrete -- the singly-charged states are
\begin{equation}\label{discrete}
\tf\gg1:\quad \frac{N_{\pm1}}{N_0}\simeq \frac{\tm^{\mp1/2}}{\tf},
\end{equation}
i.e., grains with $Z=-1$ are the most abundant; the multiply charged states ($|Z|\geq2$) are usually exponentially small and
practically negligible. We notice that the Gaussian charge distribution, usually assumed in ionization models
\citep[][]{Okuzumi2009,Fujii2011,Dzyurkevich2013,Mori2016}, may only be employed for $\tf\ll1$.

In Section~\ref{tf>>1} we identify a narrow range of $\tf~(\sim1)$ where a gradual transition between the charge triplet
(\ref{discrete}) and the Gaussian distribution (\ref{Gauss}) occurs. Also, we discuss the role of the polarization
interactions \citep[][]{Draine1987}, neglected in our consideration, and show that they have practically no effect on the
obtained results.

\section{Densities of charged species}
\label{densities}

In the dust-phase recombination regime, the self-consistent ionization degree and the grain charge distribution are
determined by a set of two equations: the ionization-recombination balance equation and the charge-neutrality equation.

\begin{table*}[!ht]\centering
\caption{Notations used in the article.}\label{tabvariab}
   \begin{tabular}{ l | l  }
     \hline
      Symbol & Meaning  \\ \hline
      $A$ & effective atomic mass of ions [amu]\\
      $a$ & dust grain radius [cm] \\
      $f_{\rm d}$ & dust-to-gas mass ratio  \\
      $m_{\rm g}, m_{\rm d}, m_{\rm e}, m_{\rm p}$ & mass of a gas particle, dust grain (of radius $a$), electron, and proton [g] \\
      $\tm$ & effective ion-to-electron mass ratio  \\
      $N_{Z}\equiv N(Z,a)$ & discrete charge distribution of grains of radius $a$ [cm$^{-4}$] \\
      $dn_{\rm d}(a)/da$ & differential size distribution of grains [cm$^{-4}$] \\
      $n_{\rm g}, n_{\rm e}, n_{\rm i}$ & number density of gas particles, electrons, and the total number density of ions [cm$^{-3}$] \\
      $n_{\rm EI}$ & number density of an electron-ion (EI) plasma [cm$^{-3}$] \\
      $n_{\rm g}^{\rm rec}$ & recombination threshold, separating the gas-phase and dust-phase recombination regimes [cm$^{-3}$] \\
      $n_{\rm g}^{\rm dep}$ & electron depletion threshold, at the transition to a dust-ion (DI) plasma [cm$^{-3}$] \\
      $n_{\rm g}^{\rm asy}$ & asymptotic threshold, at the transition to a dust-dust (DD) plasma [cm$^{-3}$] \\
      $R_{\rm g}, R_{\rm d}$ & gas-phase and dust-phase recombination rates [cm$^{-3}$~s$^{-1}$] \\
      $T$ & gas/dust temperature (same for all species) [K] \\
      $v_{\rm e}, v_{\rm i}, v_{\rm d}$ & thermal velocity scale of electrons, ions (of the atomic mass $A$),
      and grains (of radius $a$) [cm~s$^{-1}$] \\
      $Z, \langle Z\rangle$ & grain charge state and average charge number \\
      $\beta_{\rm g}$ & characteristic rate coefficient for the gas-phase recombination [cm$^{3}$~s$^{-1}$] \\
      $\beta_{\rm d}$ & rate coefficient for the dust-phase recombination [cm$^{3}$~s$^{-1}$] \\
      $\zeta$ & total ionization rate [s$^{-1}$] \\
      $\tf$ & normalized grain potential of the unit charge \\
      $\Psi=\langle Z\rangle\tf$ & normalized grain potential for charge number $\langle Z\rangle$ (limiting case $\tf\ll1$) \\
      \hline
  \end{tabular}
\end{table*}

We start with the derivation of the ionization-recombination equation. The equilibrium charge distribution discussed above
is determined by the detailed balance of the electron and ion fluxes on a grain surface, and therefore it does not matter
which of these fluxes is used to calculate the recombination. For convenience, we consider the ion collection term, which
can be presented in the following general form:
\begin{equation}\label{R_id}
R_{\rm d}=2\sqrt{2\pi}\:v_{\rm i}n_{\rm i}\int da\:a^2\mathcal{N}(a),
\end{equation}
where $v_{\rm i}=\sqrt{k_{\rm B}T/Am_{\rm p}}$ is the effective thermal velocity scale of ions and $\mathcal{N}(a)$ is the
effective number density of grains of radius $a$, obtained for the ion collection in Appendix~\ref{app_flux}. By
substituting $\mathcal{N}(a)$ from Equation~(\ref{Neff}), we derive the equation
\begin{eqnarray}
\zeta n_{\rm g}=2\sqrt{2\pi}\:v_{\rm i}n_{\rm i}\nonumber\hspace{3.5cm}\\
\times\sum_{Z\geq0}(\sqrt{\tm}+\tm^{-Z})\int da\:a^2e^{-Z\tf}N_{-Z}.\label{ion-rec}
\end{eqnarray}
Note that the sign of the index in $N_{Z}$ is inverted, so that the summation is in fact performed over {\it non-positive}
charge states ($Z\leq0$).

Next, we obtain the charge-neutrality equation. The charge density of dust grains (of a given size) is $\sum_ZZN_{Z}$, where
the summation over positive charges can be eliminated by using Equation~(\ref{distr-}). Then the integration over the size
distribution yields
\begin{equation}\label{quasi}
n_{\rm i}-n_{\rm e}=\sum_{Z>0}(1-\tm^{-Z})\int da\:ZN_{-Z},
\end{equation}
where the summation is over {\it negative} charge states.

Equations~(\ref{ion-rec}) and (\ref{quasi}) are complemented with the normalization condition for the dust density,
Equation~(\ref{norm}). Again, by eliminating the summation over positive charges we obtain
\begin{equation}\label{norm1}
N_0+\sum_{Z>0}(1+\tm^{-Z})N_{-Z}=dn_{\rm d}(a)/da.
\end{equation}
The derived set of governing Equations~(\ref{ion-rec})--(\ref{norm1}), along with the relation
\begin{equation*}
\tm=1836A(n_{\rm e}/n_{\rm i})^2
\end{equation*}
and Equation~(\ref{distr}), allows us to calculate self-consistent charge distributions for arbitrary grain-size
distributions, and obtain the corresponding electron and (total) ion densities. The grain-size distribution $dn_{\rm
d}(a)/da$, the gas density $n_{\rm g}$, and the ionization rate $\zeta$ (which is generally a decreasing function of $n_{\rm
g}$) are the input parameters for the derived model. The effective atomic mass of the ions $A$ is the only free parameter
(entering through the effective mass ratio $\tm$ and the velocity scale $v_{\rm i}\propto1/\sqrt{A}$). The model is {\it
exact} as long as the gas-phase recombination plays no role, i.e., when the strong condition (\ref{th1}) is satisfied.

The numerical solution of the derived equations, obtained for conditions of a typical protoplanetary disk and for several
characteristic grain-size distributions, is presented in Section~\ref{implications}. However, before discussing these
results, in the following Sections we study important limiting cases -- this allows us to reveal and better understand a
critical role of different physical mechanisms controlling ionization and dust charging. First, in Sections~\ref{tf<<1} and
\ref{tf>>1} we study analytically the two limiting cases of $\tf\ll1$ and $\tf\gg1$, assuming that all grains have the same
size. Then, in Section~\ref{size_effect} we analyze generic effects introduced by the size polydispersity and show how the
results derived for monodisperse particles can be generalized for an arbitrary size distribution.

For convenience, in Table~\ref{tabvariab} we summarize the main notations used throughout the paper.

\subsection{Case $\tf\ll1$ (big grains or/and high temperature)}
\label{tf<<1}

The charge distribution for $\tf\ll1$, Equation~(\ref{Gauss}), is a broad Gaussian function ($\sigma_Z\gg1$), so the
summation over $Z$ can be replaced with the integration. Moreover, as we show later, $N_Z$ in this case can be well
approximated by a shifted delta-function, $N_{Z}\propto n_{\rm d}\delta(Z-\langle Z\rangle)$ (where $n_{\rm d}$ is the
density of monodisperse particles of radius $a$; the integration over $a$ in the governing equations is removed). By
substituting the delta-function in Equations~(\ref{ion-rec}) and (\ref{quasi}), employing Equation~(\ref{Psi}), and
neglecting small terms $\tm^{-|\langle Z\rangle|}$, we reduce the governing equations to the following form:
\begin{eqnarray}
  \zeta n_{\rm g}=2\sqrt{2\pi}\:v_{\rm i}n_{\rm i}(1+\Psi)a^2n_{\rm d},\label{gov1a} \\
  n_{\rm i}-n_{\rm e} =\Psi\tf^{-1}n_{\rm d},\label{gov1b}
\end{eqnarray}
where the dust and gas densities are related by Equation~(\ref{fraction}). Together with the charging equation~(\ref{Psi}),
these equations are solved for $n_{\rm e}$, $n_{\rm i}$, and $\Psi$; they can be reduced to a single nonlinear equation for
$\Psi$, which depends on the input parameters via the ratio $n_{\rm g}/\zeta(n_{\rm g})$. The presented approach is
equivalent to that used by \citet[][]{Okuzumi2009}.

The generic behavior of the solution can be easily understood from a simple scaling analysis: As $n_{\rm d}\propto n_{\rm
g}$, from Equation~(\ref{gov1a}) it follows that $n_{\rm i}\propto\zeta$. Since $\zeta$ is generally a decreasing function
of $n_{\rm g}$, we conclude that the lhs of Equation~(\ref{gov1b}) does not increase with $n_{\rm g}$, while the rhs scales
as $\propto\Psi(n_{\rm g})n_{\rm g}$. The latter describes a contribution of the negatively charged dust to the overall
charge neutrality. As long as $n_{\rm g}$ is sufficiently small, the dust contribution is negligible and the charge
neutrality is reduced to $n_{\rm i}-n_{\rm e}\simeq0$. This corresponds to ``regular'' {\it electron-ion} (EI) plasmas,
where $\Psi=\Psi_{\rm EI}$ does not depend on $n_{\rm g}$ and is determined from Equation~(\ref{Psi}) with
$\tm=1836A\equiv\tm_{\rm EI}$. The dust contribution becomes crucial at larger $n_{\rm g}$, where Equation~(\ref{gov1b}) can
only be satisfied if $\Psi$ and, therefore, $\tm\propto (n_{\rm e}/n_{\rm i})^2$ decrease with $n_{\rm g}$, i.e., if the
electron density is depleted. Hence, there exists a certain {\it electron depletion threshold} $n_{\rm g}^{\rm dep}$ for the
gas density -- it identifies a crossover from EI to {\it dust-ion} (DI) plasmas, where the charge neutrality is regulated by
positive ions and negatively charged dust grains.

We define the electron depletion threshold $n_{\rm g}^{\rm dep}$ as $n_{\rm g}$ at which the dust charge density, given by
the rhs of Equation~(\ref{gov1b}), becomes equal to the electron density.\footnote{Mathematically, this is equivalent to the
condition that the total ion density $n_{\rm i}$, obtained from Equation~(\ref{gov1a}), is two times the dust charge
density.} This yields
\begin{equation}\label{ncr1}
\left(\frac{n_{\rm g}}{\zeta}\right)_{\rm dep}=\frac{(m_{\rm d}/m_{\rm g})^2}{4\sqrt{2\pi}f_{\rm d}^2v_{\rm i} a^2}
\frac{\tf}{\Psi_{\rm EI}(1+\Psi_{\rm EI})},
\end{equation}
where $(n_{\rm g}/\zeta)_{\rm dep}\equiv n_{\rm g}^{\rm dep}/\zeta(n_{\rm g}^{\rm dep})$ is a function of $n_{\rm g}^{\rm
dep}$. To obtain the magnitude of $n_{\rm g}^{\rm dep}$, we set for simplicity $\zeta=$~const and assume typical values used
to estimate $n_{\rm g}^{\rm rec}$ from Equation~(\ref{th1}); e.g., for $\tf\sim0.03$ this yields $n_{\rm g}^{\rm
dep}\sim10^3n_{\rm g}^{\rm rec}$. To distinguish between the gas-phase and dust-phase recombination regimes in EI plasmas,
below we adopt notations EI[g] and EI[d] for the respective density ranges (of $n_{\rm g}\lesssim n_{\rm g}^{\rm rec}$ and
$n_{\rm g}^{\rm rec}\lesssim n_{\rm g}\lesssim n_{\rm g}^{\rm dep}$).

\begin{figure*}\centering
\includegraphics[width=2\columnwidth,clip=]{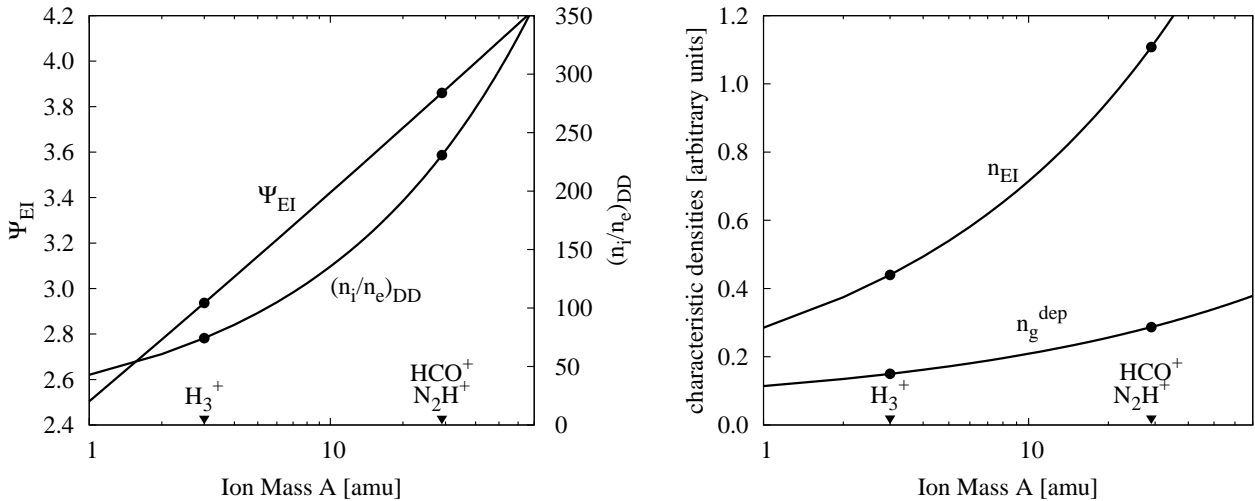}
\caption{(Left) Potential of a grain in an electron-ion (EI) plasma, $\Psi_{\rm EI}$, and the ion-to-electron density ratio
in a dust-dust (DD) plasma, $(n_{\rm i}/n_{\rm e})_{\rm DD}$, plotted as functions of the effective atomic mass of ions $A$.
(Right) Plots representing the EI plasma density, $\sqrt{A}\;(1+\Psi_{\rm EI})^{-1}\propto n_{\rm EI}(A)$, and the electron
depletion threshold, $\sqrt{A}\;\Psi_{\rm EI}^{-1}(1+\Psi_{\rm EI})^{-1}\propto n_{\rm g}^{\rm dep}(A)$; the asymptotic
threshold $n_{\rm g}^{\rm asy}(A)$ (not shown) scales as $\propto\sqrt{A}$. In both panels, values for typical ions H$_3^+$,
N$_2$H$^+$, HCO$^+$ are indicated.} \label{fig1}
\end{figure*}

\begin{figure*}\centering
\includegraphics[width=.85\textwidth,clip=]{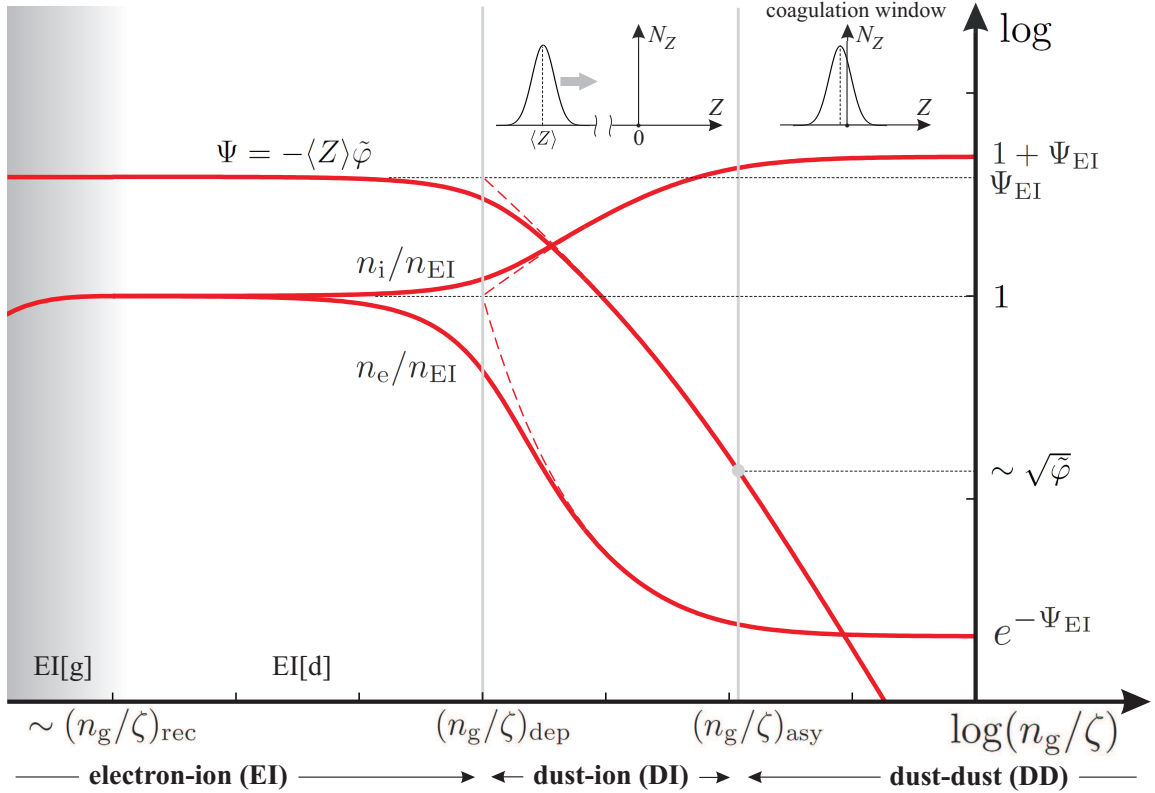}
\caption{Universal behavior of charged species in the {\it dust-phase recombination regime} $n_{\rm g}\gtrsim n_{\rm g}^{\rm
rec}$. A gradual transition, occurring in EI plasmas between the gas-phase (EI[g]) and dust-phase (EI[d]) recombination
regimes, is marked by the grey shading. The dust density is proportional to $n_{\rm g}$, all shown parameters depend on
$n_{\rm g}$ via the ratio $n_{\rm g}/\zeta(n_{\rm g})$ (the results are shown in a log-log scale, decades are indicated).
The red solid lines depict the normalized potential of a grain $\Psi$, proportional to the average dust charge $\langle
Z\rangle~<0$, as well as the normalized electron density $n_{\rm e}$ and the total ion density $n_{\rm i}$; the dashed lines
are approximate relations for $n_{\rm g}\gtrsim n_{\rm g}^{\rm dep}$. In EI plasmas, electrons and ions have the same
densities equal to $n_{\rm EI}(n_{\rm g})$, the grain potential is constant and equal to $\Psi_{\rm EI}$. A crossover to DI
plasmas occurs at the {\it electron depletion threshold} $n_{\rm g}\sim n_{\rm g}^{\rm dep}$: The ratio $n_{\rm e}/n_{\rm
i}$ and hence $\Psi$ start to decrease monotonically, while the peak of the charge distribution $N_{Z}$ moves toward $Z=0$.
The magnitude of $\langle Z\rangle$ becomes comparable to the width of $N_{Z}$ when $\Psi\sim\sqrt{\tf}~(\ll1)$, which
indicates a crossover to DD plasmas, occurring at the {\it asymptotic threshold} $n_{\rm g}\sim n_{\rm g}^{\rm asy}$:
Asymptotically, the normalized $n_{\rm i}$ slightly increases and tends to $1+\Psi_{\rm EI}$, the normalized $n_{\rm e}$
approaches a small value of $e^{-\Psi_{\rm EI}}$, while $\Psi$ tends to zero as $\propto(n_{\rm g}/\zeta)^{-1}$. The
electrostatic repulsion between charged grains virtually disappears and a ``coagulation window'' opens up (allowing a growth
of large aggregates, as discussed in Section~\ref{window}). Characteristic values of the threshold gas densities are
presented in Table~\ref{thresholds}.} \label{fig2}
\end{figure*}

In EI plasmas, the density of electrons (ions) $n_{\rm EI}$ is directly obtained from Equation~(\ref{gov1a}) with
$\Psi=\Psi_{\rm EI}$. This can be expressed in terms of $(n_{\rm g}/\zeta)_{\rm dep}$ as
\begin{equation}\label{ne1}
\frac{n_{\rm EI}}{\zeta}=2f_{\rm d}\frac{m_{\rm g}}{m_{\rm d}}\frac{\Psi_{\rm EI}}{\tf}
\left(\frac{n_{\rm g}}{\zeta}\right)_{\rm dep},
\end{equation}
i.e., the ratio $n_{\rm EI}/\zeta$ does not depend on $n_{\rm g}$ (note that it is also independent of $\tf$).
Simultaneously, this equation provides a convenient normalization for the electron and ion densities in DI plasmas. In this
case, $n_{\rm i}$ and $\Psi$ can be approximately calculated by neglecting $n_{\rm e}$ in Equation~(\ref{gov1b}). Together
with Equations~(\ref{gov1a}), (\ref{ncr1}) and (\ref{ne1}), this leads to the following simple relations for $n_{\rm
g}\gtrsim n_{\rm g}^{\rm dep}$:
\begin{eqnarray}
  \frac{\Psi(1+\Psi)}{\Psi_{\rm EI}(1+\Psi_{\rm EI})} &\simeq&\frac{(n_{\rm g}/\zeta)_{\rm dep}}{(n_{\rm g}/\zeta)},\label{Psi1} \\
  \frac{n_{\rm i}}{n_{\rm EI}} &\simeq& \frac{1+\Psi_{\rm EI}}{1+\Psi}.\label{ni1}
\end{eqnarray}
To derive $n_{\rm i}(n_{\rm g})$, one has to substitute $\Psi(n_{\rm g})$ from Equation~(\ref{Psi1}) and $n_{\rm EI}(n_{\rm
g})$ from Equation~(\ref{ne1}) in Equation~(\ref{ni1}). The electron density $n_{\rm e}$ is obtained by substituting $n_{\rm
i}$ in Equation~(\ref{Psi}), and noting that $\sqrt{\tm}=(n_{\rm e}/n_{\rm i})\sqrt{\tm_{\rm EI}}$. This yields
\begin{equation}\label{ne/ni}
\frac{n_{\rm e}}{n_{\rm EI}}\simeq e^{-(\Psi_{\rm EI}-\Psi)}.
\end{equation}
Equations~(\ref{Psi1})--(\ref{ne/ni}) show that, in a DI plasma, $\Psi$ and hence the average charge $|\langle
Z\rangle|=\tf^{-1}\Psi$ monotonically decrease with $n_{\rm g}$ and, when $\Psi\lesssim1$, tend to zero as $\Psi(n_{\rm
g})\propto\zeta/n_{\rm g}$. Correspondingly, $n_{\rm i}(n_{\rm g})$ tends to $(1+\Psi_{\rm EI})$ times $n_{\rm EI}(n_{\rm
g})$. The density ratio $n_{\rm e}/n_{\rm i}$ approaches a small (but finite) value of $1/\sqrt{\tm_{\rm EI}}$.

We remind that the results presented in this Section are obtained by substituting a shifted delta-function for the Gaussian
charge distribution (Equation~(\ref{Gauss}), entering the governing equations~(\ref{ion-rec}) and (\ref{quasi})). This
approximation is formally justified as long as the average charge $|\langle Z\rangle|=\tf^{-1}\Psi$ exceeds the charge
variance $\sigma_Z\sim\tf^{-1/2}$, i.e., for $\Psi\gtrsim\sqrt{\tf}$. Remarkably, it turns out that the derived results
remain valid also for $\Psi\lesssim \sqrt{\tf}$: Equations~(\ref{ion-rec}) and (\ref{quasi}), solved in this case with the
Gaussian distribution, directly yield relations (\ref{Psi1})--(\ref{ne/ni}).

The condition $|\langle Z\rangle|\sim\sigma_Z$ (or, equivalently, $\Psi\sim \sqrt{\tf}$) corresponds to another
characteristic gas density, the {\it asymptotic threshold} $n_{\rm g}^{\rm asy}$. Using Equation~(\ref{Psi1}), we obtain
\begin{equation}\label{ncr*}
\left(\frac{n_{\rm g}}{\zeta}\right)_{\rm asy}=\frac{\Psi_{\rm EI}(1+\Psi_{\rm EI})}{\sqrt{\tf}}
\left(\frac{n_{\rm g}}{\zeta}\right)_{\rm dep}.
\end{equation}
This threshold identifies the next important crossover, now from DI to {\it dust-dust} (DD) plasmas, where the charge
neutrality is increasingly dominated by a balance between negatively and positively charged grains. For $n_{\rm
g}/\zeta\gtrsim \tf^{-1/2}(n/\zeta)_{\rm asy}$ the average charge becomes less than unity. Since
$\sigma_Z\sim\tf^{-1/2}\gg1$ for any $n_{\rm g}$, the Gaussian charge distribution becomes asymptotically symmetric with
respect to $Z=0$.

We see that the obtained results are completely characterized by the normalized potential $\Psi_{\rm EI}$ as well as by the
characteristic densities $n_{\rm EI}$ and $n_{\rm g}^{\rm dep}$ (we notice that the asymptotic and depletion thresholds are
related by Equation~(\ref{ncr*})). All these parameters are functions of the effective atomic mass of ions $A$, which is the
only free parameter of our model. The left panel in Figure~\ref{fig1} shows $\Psi_{\rm EI}$ as well as the asymptotic
density ratio $(n_{\rm i}/n_{\rm e})_{\rm DD}$, both plotted versus $A$. In the right panel, functions representing the
dependencies $n_{\rm EI}(A)$ and $n_{\rm g}^{\rm dep}(A)$ are depicted (recall that $v_{\rm i}\propto 1/\sqrt{A}$). In
principle, $A$ may vary between unity (hydrogen ions) and some large numbers (e.g., 56 for iron ions), but astrochemical
models~\citep{Semenov2004} suggest that molecular and metal ions, such as HCO$^+$/N$_2$H$^+$ ($A=29$) and Mg$^+$ ($A=24$),
usually dominate in very dense molecular clouds and protoplanetary disks. From Figure~\ref{fig1} we see that the uncertainty
in $n_{\rm EI}$ and $n_{\rm g}^{\rm dep}$, associated with the plasma composition in this case, does not exceed a few dozens
of percent. This effect is practically negligible next to uncertainties introduced by poorly known grain-size distribution
\citep[e.g.,][]{Kim1994,Weingartner2001b} and grain morphology (the latter determines dependence of the grain mass on its
effective size).

Figure~\ref{fig2} summarizes the behavior of the average dust charge and of the electron and ion densities in the entire
dust-phase recombination regime, and also illustrates modification of the dust charge distribution with increasing $n_{\rm
g}$. The densities are normalized by $n_{\rm EI}(n_{\rm g})$, so for $n_{\rm g}\ll n_{\rm g}^{\rm dep}$ we have $n_{\rm
e}/n_{\rm EI}= n_{\rm i}/n_{\rm EI}=1$. The value of $\Psi_{\rm EI}$ weakly depends on the average atomic mass of ions $A$
(the plotted curves are for HCO$^+$/N$_2$H$^+$ ions). The solid red lines show the exact solution of the governing
Equations~(\ref{gov1a}) and (\ref{gov1b}) together with the charging Equation~(\ref{Psi}), which are equivalent to
Equations~(32)--(34) of \citet[][]{Okuzumi2009}. With the used normalization, the curves have a universal form, applicable
for arbitrary set of parameters in the dust-phase recombination regime. The behavior at $n_{\rm g}\gtrsim n_{\rm g}^{\rm
dep}$ is well reproduced by approximate relations (\ref{Psi1})--(\ref{ne/ni}), shown by the red dashed lines, with the
maximum deviation at $n_{\rm g}= n_{\rm g}^{\rm dep}$ (where the ion density is underestimated by $\simeq20\%$, while the
grain potential and the electron density are overestimated by $\simeq30\%$ and a factor of $\simeq2.5$, respectively). Note
that $\tf$ is an arbitrary small parameter in the considered case, and therefore the point $\Psi\sim\sqrt{\tf}$ (which
identifies the crossover to a DD plasma) could in principle be located below the asymptotic value of $n_{\rm e}/n_{\rm
EI}\to e^{-\Psi_{\rm EI}}$.

So far we have assumed that the gas temperature is constant. In protoplanetary disks, the relative temperature increase
toward the center is usually not as strong as the increase of the gas density. Nevertheless, this is a noticeable effect
which can in fact be straightforwardly incorporated in our model~-- the temperature simply becomes an additional input
parameter. Moreover, it turns out that in many cases, for instance~-- in the disk midplane, $T$ and $n_{\rm g}$ are related
by a simple power-law dependence, $T\propto n_{\rm g}^{\epsilon}$ \citep[e.g., with $\epsilon=2/11$, see][]{Armitage2007}.
We notice that the gas temperature enters $v_{\rm i}\propto\sqrt{T}$ and $\tf\propto T^{-1}$, thus affecting the electron
depletion threshold $n_{\rm g}^{\rm dep}$. From this we immediately infer that the ratio $(n_{\rm g}/\zeta)_{\rm dep}$ on
the lhs of Equations~(\ref{ncr1}) should be replaced with $(n_{\rm g}^{1+3\epsilon/2}/\zeta)_{\rm dep}$ (then the rhs is
properly normalized with the respective density scale entering the $T\propto n_{\rm g}^{\epsilon}$ dependence); both ratios
on the rhs of Equation~(\ref{Psi1}) are replaced in the same way. As regards $n_{\rm EI}$, the lhs of Equation~(\ref{ne1})
should be multiplied with $(n_{\rm g}/n_{\rm g}^{\rm dep})^{\epsilon/2}$, i.e., $n_{\rm EI}$ falls off with gas density as
$\propto\zeta/n_{\rm g}^{\epsilon/2}$.

When considering recombination on grains, we have so far also implicitly assumed that only free electrons and ions
contribute to this process. Such an approach is natural since the thermal velocities of plasma species are much larger than
typical thermal velocities of grains, and therefore the terms describing recombination due to mutual dust collisions
\citep[][]{Umebayashi1983,Umebayashi1990,Marchand2016} are omitted in Equation~(\ref{ion-rec}). On the other hand, in a DD
plasma grains become the most abundant charged species and, thus, collisions between them may provide an important
contribution to the net recombination rate. This occurs when the product $n_{\rm d}v_{\rm d}$ becomes comparable to the
thermal ion flux $n_{\rm i}v_{\rm i}$, where $v_{\rm d}$ is the relevant (thermal or non-thermal) scale for the relative
velocity of grains. By substituting the asymptotic expression for $n_{\rm i}(n_{\rm g})$ we conclude that the recombination
mechanism due to mutual dust collisions should be important at $n_{\rm g}/\zeta\gtrsim (v_{\rm i}/v_{\rm d}) (n_{\rm
g}/\zeta)_{\rm dep}$. If the thermal motion dominates dust dynamics, the corresponding gas density exceeds $n_{\rm g}^{\rm
dep}$ by many orders of magnitudes: e.g., for micron-size grains in a HCO$^+$ plasma, $n_{\rm g}$ should be of the order of
$3\times10^5n_{\rm g}^{\rm dep}~(\sim10^4 n_{\rm g}^{\rm asy}$). However, the condition can be significantly relaxed for
grains exhibiting strong non-thermal motion, e.g., due to differential drift or sedimentation \citep[][]{Testi2014}.
Furthermore, since $(n_{\rm g}/\zeta)_{\rm dep}\propto f_{\rm d}^{-2}$, an increase of the dust fraction occurring due to
various processes operating in the disk midplane (see Section~\ref{window}) also promotes this mechanism of recombination.

\subsection{Case $\tf\gg1$ (small grains or/and low temperature)}
\label{tf>>1}

In this case the charge distribution is very different from the Gaussian form -- practically, it is limited by the
singly-charged and neutral states, as it follows from Equation~(\ref{discrete}). By substituting the charge distribution in
Equation~(\ref{norm1}) we derive
\begin{eqnarray}
  N_0\simeq(1+\sqrt{\tm}/\tf\:)^{-1}n_{\rm d}, \label{N0}\\
  N_{-1}\simeq(1+\tf/\sqrt{\tm}\:)^{-1}n_{\rm d}, \label{N-1}
\end{eqnarray}
where small terms $1/(\sqrt{\tm}\tf)$ were neglected. The governing Equations~(\ref{ion-rec}) and (\ref{quasi}) are reduced
to
\begin{eqnarray}
  \zeta n_{\rm g} = 2\sqrt{2\pi}\:v_{\rm i}n_{\rm i}(1+\sqrt{\tm}\:)a^2N_0,\label{gov2a} \\
  n_{\rm i}-n_{\rm e} = (1-\tm^{-1})N_{-1},\label{gov2b}
\end{eqnarray}
where $\tm=(n_{\rm e}/n_{\rm i})^2\tm_{\rm EI}$. By substituting Equation~(\ref{N-1}) in Equation~(\ref{gov2b}) we obtain
the average charge
\begin{equation}\label{2_mch}
|\langle Z\rangle|=\frac{1-\tm^{-1}}{1+\tf/\sqrt{\tm}},
\end{equation}
which is always smaller than unity and, as expected, tends to zero when $\tm\to1$.

For an EI plasma one should set $\tm=\tm_{\rm EI}$ and neglect the rhs of the charge-neutrality Equation~(\ref{gov2b}),
exactly as in the case $\tf\ll1$. The corresponding electron depletion threshold is determined by
\begin{equation}\label{ncr2}
\left(\frac{n_{\rm g}}{\zeta}\right)_{\rm dep}=\frac{(m_{\rm d}/m_{\rm g})^2}{4\sqrt{2\pi}f_{\rm d}^2v_{\rm i} a^2}
\frac{(1+\tf/\sqrt{\tm_{\rm EI}}\:)^2}{\tf},
\end{equation}
and the plasma density for $n_{\rm g}\lesssim n_{\rm g}^{\rm dep}$ is
\begin{equation}\label{ne2}
\frac{n_{\rm EI}}{\zeta}=2f_{\rm d}\frac{m_{\rm g}}{m_{\rm d}}(1+\tf/\sqrt{\tm_{\rm EI}}\:)^{-1}
\left(\frac{n_{\rm g}}{\zeta}\right)_{\rm dep}.
\end{equation}
We take into account that $\tm_{\rm EI}$ is very large, so the terms $\sim\tm_{\rm EI}^{-1}$ and $\sim1/\sqrt{\tm_{\rm EI}}$
are omitted in both expressions (whereas the retained terms $\tf/\sqrt{\tm_{\rm EI}}$ may be arbitrary large in the
considered case).

For a DI plasma, $n_{\rm g}\gtrsim n_{\rm g}^{\rm dep}$, we derive the following relations:
\begin{eqnarray}
(1-\tm^{-1})(1+1/\sqrt{\tm}\:)\left(\frac{1+\tf/\sqrt{\tm_{\rm EI}}}{1+\tf/\sqrt{\tm}}\right)^2\hspace{1cm}\nonumber\\
\simeq\frac{(n_{\rm g}/\zeta)_{\rm dep}}{(n_{\rm g}/\zeta)}, \label{tm}\\
\frac{n_{\rm i}}{n_{\rm EI}} \simeq (1+1/\sqrt{\tm}\:)^{-1}\left(\frac{1+\tf/\sqrt{\tm}}{1+\tf/\sqrt{\tm_{\rm EI}}}\right).
\label{ni2}
\end{eqnarray}
Equation~(\ref{tm}) yields the solution for $\tm(n_{\rm g})$ and, hence, for $n_{\rm e}/n_{\rm i}=\sqrt{\tm/\tm_{\rm EI}}$;
the dependence $n_{\rm i}(n_{\rm g})$ is obtained by substituting $\tm(n_{\rm g})$ in Equation~(\ref{ni2}). Asymptotically
we obtain $\tm-1\propto|\langle Z\rangle|\propto\zeta/n_{\rm g}$ and $n_{\rm i}\propto\zeta$.

Thus, the qualitative behavior of $n_{\rm e}$, $n_{\rm i}$, and $|\langle Z\rangle|$ remains exactly the same as in the case
$\tf\ll1$. On the other hand, the values of $n_{\rm g}^{\rm dep}$ (Equations~(\ref{ncr1}) or (\ref{ncr2})) and $n_{\rm EI}$
(Equations~(\ref{ne1}) or (\ref{ne2})) can be quite different in the two cases -- their relative magnitudes depend on
$\Psi_{\rm EI}$ and $\tf$ (for the corresponding case). Nevertheless, the curves for the normalized electron and ion
densities, plotted versus the normalized gas density in the case $\tf\gg1$, look similar to those in Figure~\ref{fig2}.
Interestingly, the absolute value of the asymptotic ion density $n_{\rm i}(n_{\rm g})$ for $\tf\gg1$ is exactly a half of
that derived for $\tf\ll1$. The dust charge distribution gradually changes at $n_{\rm g}\gtrsim n_{\rm g}^{\rm dep}$, from
the asymmetric triplet with the maximum at $Z=-1$, as given by Equation~(\ref{discrete}) with $\tm=\tm_{\rm EI}$, to a
quasi-symmetric triplet $N_{\pm1}/N_0\simeq \tf^{-1}$ with the peak at $Z=0$. This latter asymptotic form represents a DD
plasma discussed in Section~\ref{tf<<1}.

Above, we have completely neglected the polarization interactions of electrons and ions with dust grains. These interactions
noticeably increase the electron/ion collection cross sections by uncharged (or weakly charged) grains in the case $\tf\gg1$
\citep[][]{Draine1987}. As a result, the relative abundances $N_{\pm1}/N_0$ in Equation~(\ref{discrete}) are increased by
the factor of $\simeq\sqrt{\pi\tf/8}$. Correspondingly, $\sqrt{\tm}$ in Equations~(\ref{N0}) and (\ref{N-1}) should be
multiplied with this factor, while $\sqrt{\tm}$ in Equation~(\ref{gov2a}) should be multiplied with $\sqrt{\pi\tf/2}$, and
Equation~(\ref{gov2b}) is left unchanged. In practice, such modification does not affect the characteristic densities given
by Equations~(\ref{ncr2}) and (\ref{ne2}), as this becomes important only for extremely large values of
$\tf\gtrsim\tm$~($\sim5\times10^4$ for a HCO$^+$/N$_2$H$^+$ plasma).

To conclude this Section, let us make a note on a crossover between the limiting cases $\tf\ll1$ and $\tf\gg1$. In an EI
plasma, the magnitude of the average charge given by Equation~(\ref{2_mch}) is always significantly smaller than $|\langle
Z_{\rm EI}\rangle|~(=\Psi_{\rm EI}/\tf$) for $\tf\ll1$. This fact does not allow us to smoothly match these two cases: From
Equations~(\ref{ncr1}) and (\ref{ncr2}) we conclude that the value of $(n_{\rm g}/\zeta)_{\rm dep}$ for $\tf\gg1$ is a
factor of $\Psi_{\rm EI}(1+\Psi_{\rm EI})$ larger than that for $\tf\ll1$, when compared at the formal ``matching point''
$\tf=1$. Similarly, from Equations~(\ref{ne1}) and (\ref{ne2}) we obtain that $n_{\rm EI}/\zeta$ calculated at the matching
point for $\tf\gg1$ is $1+\Psi_{\rm EI}$ times the value for $\tf\ll1$. Setting for simplicity $\zeta=$~const and taking a
HCO$^+$/N$_2$H$^+$ plasma as an example, we obtain the relative mismatch of $\simeq19$ for $n_{\rm g}^{\rm dep}$ and
$\simeq4.8$ for $n_{\rm EI}$. The reason for the discrepancy is obvious: Equations~(\ref{distr+}) and (\ref{distr-}) imply
that the negatively charged states with $|Z|\geq2$, neglected in Equations~(\ref{gov2a}) and (\ref{gov2b}), rapidly become
dominant when $\tf$ decreases below a value of $\simeq\ln\sqrt{\tm_{\rm EI}}$~($\simeq5.4$ for a HCO$^+$/N$_2$H$^+$ plasma).
Hence, there is a relatively narrow range (of $1\lesssim\tf\lesssim\ln\sqrt{\tm_{\rm EI}}$) where a gradual transition
between the triplet (\ref{discrete}) and the Gaussian distribution (\ref{Gauss}) takes place.

\section{Effect of grain-size distribution}
\label{size_effect}

The grain polydispersity, i.e., the presence of grains of different sizes, plays an exceptionally important role in the
discussed processes. The reason for that is twofold, as one can directly see from the governing equations: (i) The integral
in the ionization-recombination Equation~(\ref{ion-rec}) determines the magnitude of the plasma recombination rate on the
grain surface and, hence, the equilibrium plasma (ion) density. The integral can be dominated by different parts of the size
distribution, depending on its particular form. (ii) In the same way, the size distribution affects the integral determining
the contribution of charged grains in the charge-neutrality Equation~(\ref{quasi}).

Let us first consider the case $\tf\ll1$ which, remarkably, allows us to obtain {\it rigorous} results for arbitrary size
distribution from the solution derived in Section~\ref{tf<<1} for single-size grains. Indeed, keeping the integrals in
Equations~(\ref{ion-rec}) and (\ref{quasi}) yields in this case Equations~(\ref{gov1a}) and (\ref{gov1b}) with the modified
rhs: The ``recombination factor'' $a^2n_{\rm d}$ in the former equation is replaced with $\int dn_{\rm d}\:a^2$, and,
similarly, the ``charge-neutrality factor'' $\tf^{-1}n_{\rm d}$ in the latter equation is replaced with $\int dn_{\rm
d}\:\tf^{-1}$ (we remind that $\tf^{-1}\propto a$). Thus, the effects of polydispersity are {\it factorized}, leading to a
simple renormalization of the characteristic densities.

For illustration, we employ a widely used power-law dependence for the differential dust density,
\begin{equation}\label{size_distr}
dn_{\rm d}(a)/da=Ca^{-p},
\end{equation}
defined for the size range $a_{\rm min}\leq a\leq a_{\rm max}$ (here, we naturally assume that the condition $\tf\ll1$ is
satisfied for all sizes down to $a_{\rm min}$). The constant $C$ is determined from the dust-to-gas density ratio,
Equation~(\ref{fraction}), where $m_{\rm d}n_{\rm d}$ should be replaced by the integral $\int dn_{\rm d}\:m_{\rm d}$. For
certainty, we assume that the power-law index $p$ does not exceed the MRN value of 3.5 \citep[][]{Kim1994,Weingartner2001b},
so the mass integral is always dominated by the upper-size cutoff $a_{\rm max}$. For this size distribution, the integral
$\int dn_{\rm d}\:a^2$ in the modified Equations~(\ref{gov1a}) is equal to $a_{\rm max}^2n_{\rm d}$ multiplied by the
following renormalization factor:
\begin{equation}\label{replace1}
\frac{4-p}{|3-p|}\left\{
\begin{array}{cc}
1, & p<3;\\
\tilde a^{p-3}, & p>3.
\end{array}
\right.
\end{equation}
Here, $\tilde a\equiv a_{\rm max}/a_{\rm min}\gg1$ and $n_{\rm d}$ is the effective dust density determined by the condition
$\int dn_{\rm d}\: m_{\rm d}=m_{d,\rm max}n_{\rm d}$, with $m_{d,\rm max}\equiv m_{\rm d}(a_{\rm max})$. For the sake of
clarity we do not consider situations with $p\simeq3$ (where the contributions of the upper and lower cutoffs are
comparable), which allows us to neglect small terms $\propto\tilde a^{-|p-3|}$. Similarly, the integral $\tf^{-1}n_{\rm d}$
in the modified Equation~(\ref{gov1b}) is equal to $\tf_{\rm max}^{-1}n_{\rm d}$ multiplied by the renormalization factor
\begin{equation}\label{replace2}
\frac{4-p}{|2-p|}\left\{
\begin{array}{cc}
1, & p<2;\\
\tilde a^{p-2}, & p>2,
\end{array}
\right.
\end{equation}
where $\tf_{\rm max}\equiv\tf(a_{\rm max})$. We see that the contribution of smaller grains can become dominant when
$dn_{\rm d}(a)/da$ decreases sufficiently steeply with size. This lowers the plasma density (due to enhanced recombination) and
increases the average grain charge, as described by Equations~(\ref{replace1}) and (\ref{replace2}), respectively.

>From Equations~(\ref{replace1}) and (\ref{replace2}) we infer that the characteristic densities can be dramatically
decreased for polydisperse grains: From Equation~(\ref{th1}) we conclude that the resulting recombination threshold $(n_{\rm
g}/\zeta)_{\rm rec}$ is inversely proportional to the {\it squared} dust-phase recombination rate and, hence, is reduced by
the {\it squared} factor~(\ref{replace1}). For the MRN distribution this implies a decrease by a factor of about $a_{\rm
max}/a_{\rm min}\simeq50$ \citep[][]{Mathis1977}. As for the electron depletion threshold $(n_{\rm g}/\zeta)_{\rm dep}$, its
value is inversely proportional to the product of the two factors, i.e., the rhs of Equation~(\ref{ncr1}) should be divided
by
\begin{equation}\label{replace3}
\frac{(4-p)^2}{|(2-p)(3-p)|}\left\{
\begin{array}{cc}
1, & p<2;\\
\tilde a^{p-2}, & 2<p<3;\\
\tilde a^{2p-5}, & p>3.
\end{array}
\right.
\end{equation}
Again, taking the MRN distribution as an extreme example, we obtain a reduction by almost three orders of magnitude!
Finally, the density of an EI plasma, Equation~(\ref{ne1}), is directly derived from Equation~(\ref{gov1a}). Therefore,
$n_{\rm EI}/\zeta$ for polydisperse grains should be divided by the factor~(\ref{replace1}).

In the case $\tf\gg1$, the effects of polydispersity are generally no longer factorized: Now, the factors $a^2N_0$ and
$N_{-1}$ on the rhs of Equations~(\ref{gov2a}) and (\ref{gov2b}), respectively, should be replaced with integrals. Using
Equations~(\ref{N0}) and (\ref{N-1}) we obtain the respective integrals, $\int dn_{\rm d}\:a^2(1+\sqrt{\tm}/\tf)^{-1}$ and
$\int dn_{\rm d}\: (1+\tf/\sqrt{\tm})^{-1}$. One can see that a renormalization of the characteristic densities is only
possible when the range of $\tf$ (corresponding to a given range of grain sizes) does not overlap with the value of
$\sqrt{\tm}$. As the latter decreases monotonically with $n_{\rm g}$, from $\sqrt{\tm_{\rm EI}}~(=231$ for a
HCO$^+$/N$_2$H$^+$ plasma) to unity, such situation appears unlikely, and thus the governing equations should be solved
numerically. Nevertheless, in the following Section (where the numerical results are presented) we demonstrate that the
qualitative effects of polydispersity remain similar to those discussed above for the case $\tf\ll1$.

\begin{table}[!ht]\centering
\caption{Characteristic values of the threshold gas densities for $\zeta=10^{-17}$~s$^{-1}$, $T=100$~K, and two ``standard''
models of the grain-size distribution.} \label{thresholds}
   \begin{tabular}{ l | c | c | c }
     \hline
      Grain-size distribution & $n_{\rm g}^{\rm rec}$,~cm$^{-3}$ & $n_{\rm g}^{\rm dep}$,~cm$^{-3}$ & $n_{\rm g}^{\rm asy}$,~cm$^{-3}$ \\ \hline
      MRN & $2\times10^5$ & $3\times10^6$ & $1\times10^9$ \\
      $a=0.1~\mu$m & $8\times10^6$ & $2\times10^9$ & $2\times10^{11}$ \\
      \hline
  \end{tabular}
\end{table}

For convenience, in Table~\ref{thresholds} we summarize characteristic values of the threshold gas densities. Using these
values and taking into account the following scaling dependencies (for compact grains): $n_{\rm g}^{\rm rec}\propto \zeta
a^2/T$, $n_{\rm g}^{\rm dep}\propto \zeta a^3/T^{3/2}$, and $n_{\rm g}^{\rm asy}\propto \zeta a^{7/2}/T$, one can deduce the
thresholds for arbitrary parameters.

\section{Implications for Protoplanetary Disks}
\label{implications}

In this Section we employ a typical protoplanetary disk model to numerically calculate the gas ionization and dust charging.
In particular, the results are aimed to answer the following questions:
\begin{itemize}
 \item  May the ``coagulation window'' be opened, to overcome the electrostatic barrier?
 \item  How accurate is the (conventional) assumption of low dust charges?
\end{itemize}

We restrict ourselves by considering the disk midplane region, where the gas and dust mass is concentrated. The gas surface
density of the disk is assumed to obey a power-law dependence on the disk radius, $\Sigma(R)=\Sigma_{\rm AU}(R/{\rm
AU})^{-q}$ with $\Sigma_{\rm AU}=200$~g/cm$^2$ and $q=1$, which leads to the disk mass of $0.02~M_{\rm \odot}$ ($0.04~M_{\rm
\odot})$ for the outer radius of $150$~AU ($300$~AU). We consider a sharp cutoff, as a tapered power-law profile does not
qualitatively change the results. Furthermore, we assume that the disk thermal structure is determined by re-radiation of
the central star emission. The gas/dust temperature is calculated from $T^4(R)=T_*^4(R_*/R)^2\sin\phi$ \citep{Brauer2008}
with the effective star temperature $T_*=4000$~K, stellar radius $R_*=2.6~R_{\rm \odot}$, and $\sin\phi =0.05$ for the sine
of the grazing angle.\footnote{The additional gas heating due to accretion may increase the inner disk temperature, leading
to thermal ionization. Here we neglect this effect, since it is only important at high temperatures above $\sim1000$~K.} The
disk is assumed to be vertically isothermal, with the vertical density structure derived from the hydrostatic equilibrium
for the central star mass of $M_*=0.7~M_{\rm \odot}$. We consider CRs, stellar X-rays and radionuclides as primary sources
of ionization in the midplane \citep{Turner2008},
\begin{equation}\label{ionform}
 \zeta(R)=\zeta_{\rm CR}e^{-\frac{\Sigma(R)}{2\Sigma_{\rm CR}}}+\frac{\zeta_{\rm XR}}{(R/{\rm AU})^2}
 e^{-\frac{\Sigma(R)}{2\Sigma_{\rm XR}}}+\zeta_{\rm RA},
\end{equation}
where $\zeta_{\rm CR}=10^{-17}$~s$^{-1}$ is the (unattenuated) CR-ionization rate with the attenuation surface density of
$\Sigma_{\rm CR}=96$~g/cm$^2$, $\zeta_{\rm XR}=5.2\times10^{-15}$~s$^{-1}$ is the X-ray ionization rate at 1~AU (which
corresponds to the X-ray luminosity of the central star of $\simeq2\times 10^{30}$~erg/s) with the attenuation surface
density of $\Sigma_{\rm XR}=8$~g/cm$^2$, and $\zeta_{\rm RA}=10^{-21}$~s$^{-1}$ is the ionization rate due to long-lived
radioactive nuclei.

An important ingredient of any disk model are dust properties, i.e., the dust density, size distribution, and grain
morphology. Below we consider four characteristic models for the dust size distribution -- two ``monodisperse''
(single-size) populations of grains, and two MRN-like distributions \citep{Mathis1977}, given by Equation~(\ref{size_distr})
with $p=3.5$:
\begin{itemize}
\item monodisperse, $a=0.1~\mu$m;
\item monodisperse, $a=10~\mu$m;
\item MRN: $a_{\rm min}=0.005~\mu$m, $a_{\rm max}=0.25~\mu$m;
\item evolved MRN: $a_{\rm min}=0.5~\mu$m, $a_{\rm max}=25~\mu$m.
\end{itemize}
For all models, grains are supposed to be compact spheres, and the dust-to-gas ratio is the same and equal to $f_{\rm
d}=0.01$ at any location in the disk. We adopt 3.5~g/cm$^3$ for the solid mass density of dust grains, and assume that
N$_2$H$^+$ or HCO$^+$ are the dominant ions ($A=29$).

\subsection{May the ``coagulation window'' be opened?}
\label{window}

\begin{figure*}\centering
\includegraphics[width=1.8\columnwidth,clip=]{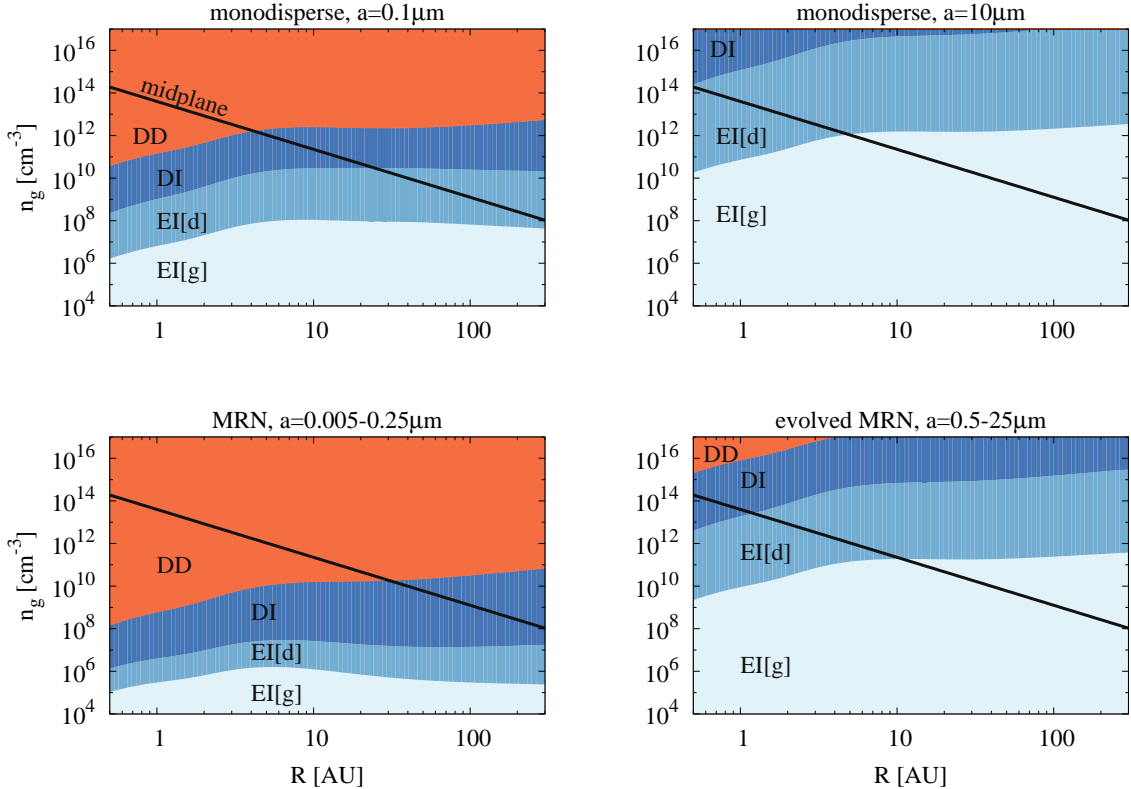}
\caption{Gas number density $n_{\rm g}$ in the disk midplane versus the radial distance $R$ (black solid line), plotted for
the assumed disk model. Different panels represent different models for the dust size distribution (as indicated, see text
for details). Boundaries between different regions in each panel are the thresholds $\left(n_{\rm g}/\zeta\right)_{\rm
rec}$, $\left(n_{\rm g}/\zeta\right)_{\rm dep}$, and $\left(n_{\rm g}/\zeta\right)_{\rm asy}$ plotted as functions of $R$
and separating, respectively, electron-ion plasmas with the gas-phase (EI[g]) and dust-phase (EI[d]) recombination, dust-ion
plasmas (DI), and dust-dust plasmas (DD).} \label{fig3}
\end{figure*}

The electrostatic barrier against the dust growth, caused by the mutual repulsion of the negatively charged grains
\citep{Okuzumi2009}, presents a fundamental but still poorly investigated issue of the modern dust evolution models. The
issue stems from estimates \citep[][]{Okuzumi2009} showing that, in most regions of a protoplanetary disk, the average
electrostatic energy of interaction between grains at their contact can exceed their relative kinetic energy. Indeed, for
``big'' grains ($\tf\ll1$) in EI plasmas, the ratio of the electrostatic energy to the thermal energy, $\simeq
\frac12|\langle Z\rangle|\Psi_{\rm EI}$, is always very large.

The two obvious ways to overcome the electrostatic barrier are to decrease the magnitude of the grain charges or to increase
the relative velocity of the grains. Recently it was shown that the photoelectric emission caused by stellar UV radiation
may drive grain charges to positive values and thus allow dust coagulation at intermediate heights of the protoplanetary
disks \citep{Akimkin2015}; similarly, the photoelectric grain charging due to CR-induced H$_2$ fluorescence can operate in
the much deeper, outer midplane regions of the disk \citep{Ivlev2015b}. None of these mechanisms, however, is able to affect
dust charging in dense disk regions. \citet{Okuzumi2011a} suggested that the presence of a large number of small grains may
remove free electrons from the gas in these regions, and thus make larger grains less charged. Let us elaborate on the
latter mechanism.

Under low-density/high-ionization conditions with $n_{\rm g}/\zeta\lesssim(n_{\rm g}/\zeta)_{\rm dep}$ (EI plasmas), grains
are highly charged due to abundance of free electrons. The average grain charge in this case tends to the maximum possible
value of $\tf^{-1}\Psi_{\rm EI}$, determined by the ion mass and temperature, and therefore the coagulation of micron-size
(or larger) grains is usually hampered. In denser regions of the disk, where $(n_{\rm g}/\zeta)_{\rm dep}<n_{\rm
g}/\zeta<(n_{\rm g}/\zeta)_{\rm asy}$ (DI plasmas), the grain charges lower due to depletion of free electrons. A specific
feature of DI plasmas is that charging of grains of a given size is determined by the entire dust ensemble. By moving into
the densest disk regions, where $n_{\rm g}/\zeta>(n_{\rm g}/\zeta)_{\rm asy}$ (DD plasmas), the depletion of electrons
eventually becomes so strong that their accretion onto a neutral grain is practically equal to the ion accretion. Thus, the
grain charge distribution becomes practically symmetric with respect to zero, opening up the opportunity for barrier-free
coagulation.

The threshold parameters $(n_{\rm g}/\zeta)_{\rm rec}$, $(n_{\rm g}/\zeta)_{\rm dep}$, and $(n_{\rm g}/\zeta)_{\rm asy}$,
identifying boundaries between different plasma regions, vary across the disk. As introduced above, $(n_{\rm g}/\zeta)_{\rm
rec}$ is determined by equal gas and dust contributions to the recombination rates, $(n_{\rm g}/\zeta)_{\rm dep}$
corresponds to equal number densities of free electrons and electrons carried by dust grains, while at $(n_{\rm
g}/\zeta)_{\rm asy}$ the total positive charge is equally distributed between free ions and grains. We express these
parameters in terms of the gas density, with the ionization rate according to Equation~\eqref{ionform}, and depict the
resulting plasma regions EI[g], EI[d], DI, and DD in Figure~\ref{fig3} for the four dust models.

Figure~\ref{fig3} shows that for the monodisperse $0.1~\mu$m dust model (upper left panel), the DD plasma region is limited
to $R\lesssim4$~AU and the EI region is at $R\gtrsim20$~AU (with the DI region filling the gap between them). Introducing
the grain polydispersity leads to a strong shift of the plasma boundaries: for the MRN model (lower left panel), the DD
plasma region extends up to $R\sim30$~AU, while the DI region continues beyond $R\sim300$~AU. These results confirm
conclusions of Section~\ref{size_effect}, demonstrating that excess of small particles in a broad size distribution may
dramatically reduce the values of $(n_{\rm g}/\zeta)_{\rm dep}$ and $(n_{\rm g}/\zeta)_{\rm asy}$. The increase of the
overall grain size has the opposite effect: According to the results of Sections~\ref{tf<<1} and \ref{tf>>1}, the boundaries
between different plasma regions in this case are shifted to much higher $n_{\rm g}$ (i.e., to smaller $R$), with the
strongest effect being on $(n_{\rm g}/\zeta)_{\rm dep}\propto a^4$ (for monodisperse grains). Indeed, the right panels in
Figure~\ref{fig3}, depicting the results for monodisperse $10~\mu$m grains and for the evolved MRN model, show no presence
of a DD plasma.

Thus, the (initial) MRN size distribution ensures that the coagulation window is opened (DD plasma state) in the entire
inner disk, promoting rapid dust growth. However, the latter process rapidly modifies the overall charge balance towards DI
and EI plasmas: for the evolved dust distributions (only by two orders of magnitude in size, as in the right panels of
Figure~\ref{fig3}) the electrostatic barrier is completely restored (EI plasma state) for $R\gtrsim1$~AU. The excess of
small grains makes the coagulation conditions more favorable, but this factor alone is insufficient for further efficient
dust growth.

We conclude that some feedbacks are necessary to facilitate the coagulation. This could be dust fragmentation leading to,
e.g., a bimodal small/big size distribution \citep[where small grains provide conditions for a reduced potential barrier
between big grains,][]{Okuzumi2011a}; or this could be an increase in the dust-to-gas ratio $f_{\rm d}$
\citep[][]{Booth2016,Surville2016}, since both the EI-DI and DI-DD plasma boundaries are determined by the value of $(n_{\rm
g}/\zeta)_{\rm dep}\propto f_{\rm d}^{-2}$. Such feedbacks may be achieved due to turbulent motion outside of the dead zone
(see Section~\ref{discussion}) -- which simultaneously increases the relative grain velocities \citep[][]{Testi2014} and
hence further reduces the effect of the electrostatic barrier. Turbulence generates local dust traps
\citep{vanderMarel2013,Flock2015,Ruge2016} where $f_{\rm d}$ can become as high as a few \citep{Johansen2007,Surville2016}.
With such values of $f_{\rm d}$, the coagulation window opens up at $R\sim1$~AU even for the monodisperse $10~\mu$m dust
(while for the evolved MRN, $f_{\rm d}\simeq0.2$ is sufficient).

\begin{figure*}\centering
\includegraphics[width=2.0\columnwidth,clip=]{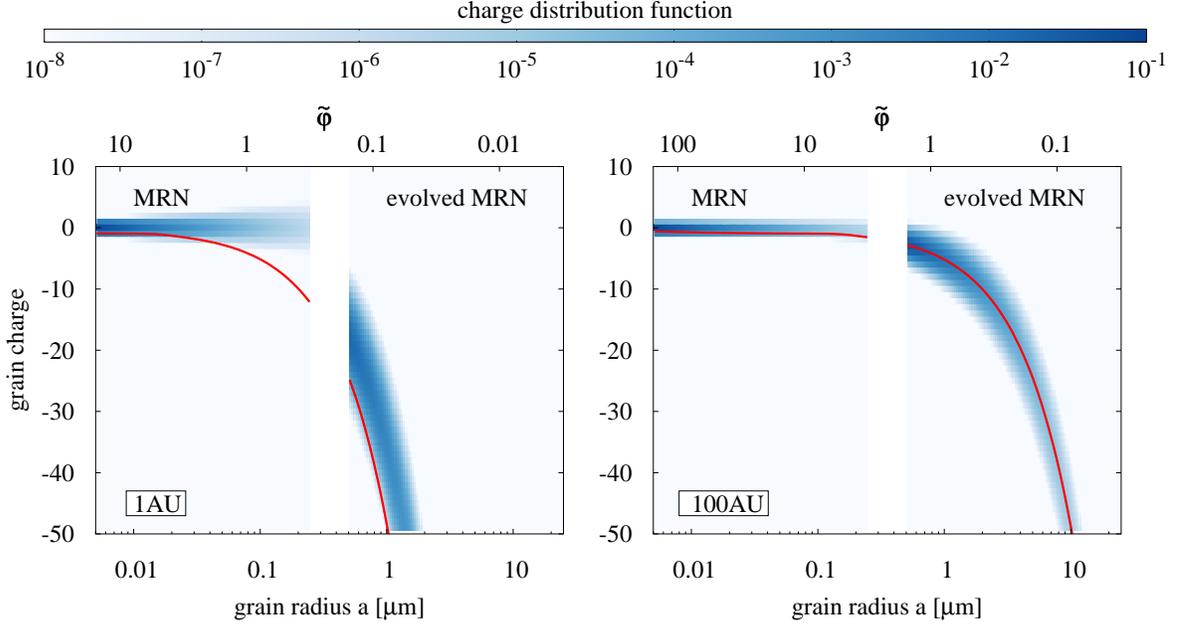}
\caption{Grain charge distribution $N(Z,a)$ for different regions in the disk midplane. The results are for two
characteristic dust size distributions: MRN $[0.005~\mu{\rm m},\;0.25~\mu{\rm m}]$ and evolved MRN $[0.5~\mu{\rm
m},\;25~\mu{\rm m}]$, the upper horizontal axis shows the corresponding values of $\tf$. The blue scale denotes the relative
number density of charged grains ($N_Z$ normalized by the total dust density $n_{\rm d}$) at the radial distance of $R=1$~AU
(left) and 100~AU (right); the respective physical conditions are $T=208$~K, $n_{\rm g}=4\times10^{13}~{\rm cm}^{-3}$,
$\zeta=4\times10^{-18}~{\rm s}^{-1}$ (left) and $T=21$~K, $n_{\rm g}=10^{9}~{\rm cm}^{-3}$, $\zeta=10^{-17}~{\rm s}^{-1}$
(right). The red solid line shows the mean grain charge calculated neglecting the charge depletion effects (i.e., assuming
$n_{\rm e}= n_{\rm i}$).} \label{fig4}
\end{figure*}

\subsection{Low charges for dust grains: Is this always justified?}
\label{singlecharge}

\begin{figure*}\centering
\includegraphics[width=1.8\columnwidth,clip=]{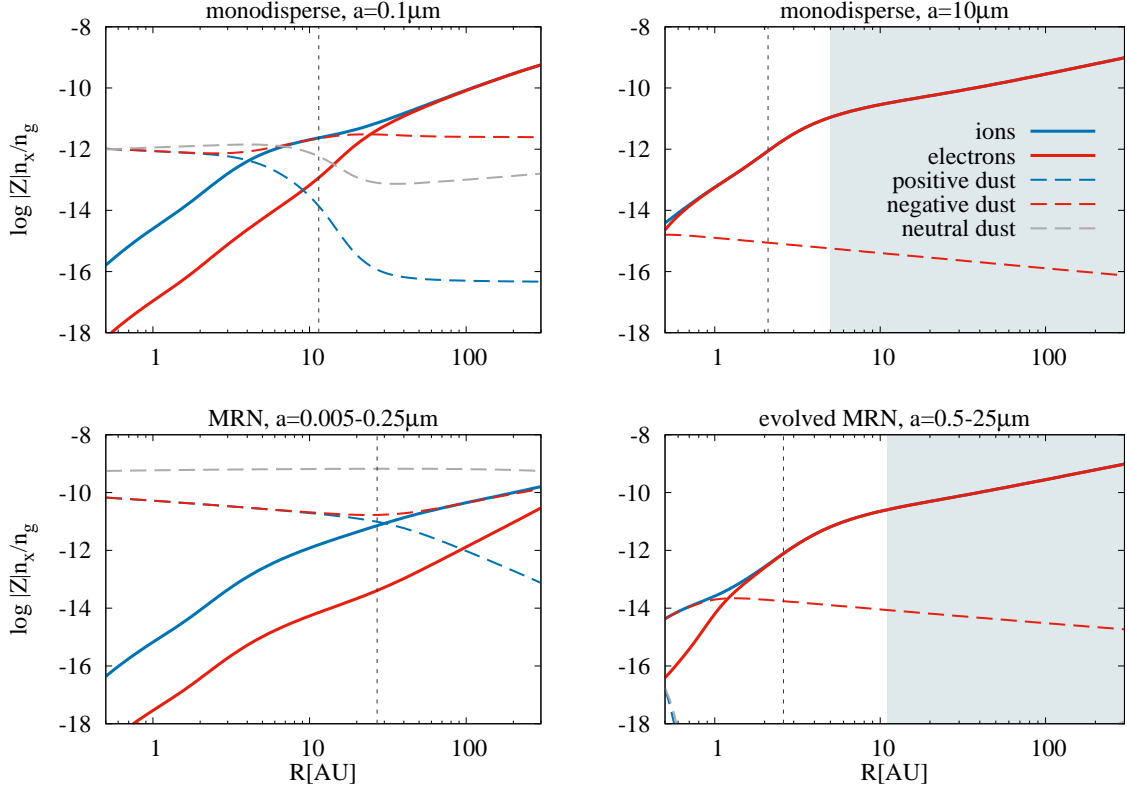}
\caption{Relative ``charge abundance'' of different species in the disk midplane, $|Z|n_{\rm x}/n_{\rm g}$, versus the
radial distance $R$. The abundances, plotted for different models of the dust size distribution (as indicated), are defined
as $n_{\rm i,e}/n_{\rm g}$ for ions and electrons, $\sum_{Z>0}\int da\:ZN_Z/n_{\rm g}$ for positive dust, $\sum_{Z<0}\int
da\:|Z|N_Z/n_{\rm g}$ for negative dust, and $\int da\:N_0/n_{\rm g}$ for neutral dust. The vertical dashed line in each
panel indicates the outer boundary of the dead zone. The shaded regions show EI[g] plasmas where the results become
inaccurate.} \label{fig5}
\end{figure*}

In the contemporary MRI and astrochemical models of protoplanetary disks, it is routinely assumed that the charge states of
dust grains are around zero ($0,\pm1,\pm2$). This is generally true for small dust grains ($a\lesssim0.1~\mu$m) in
low-temperature gas ($T\lesssim100$~K) \citep[see Equation~(5.75)]{Tielens2005}. However, such small grains cannot be
representative for the dense protoplanetary environments. As the average charge of larger grains scales linearly with the
size, the coagulation inevitably leads to the breakdown of the low-charge assumption \cite[see
also][]{Perez-Becker2011,Ilgner2012}.

In Figure~\ref{fig4} we demonstrate the grain charge distributions for the MRN and evolved MRN models, plotted at the radial
distance of 1~AU (left) and 100~AU (right). Note that different dust models correspond to independent simulations and are
only shown on the same plot to facilitate the comparison.

In the MRN case shown in Figure~\ref{fig4} most grains, indeed, carry low charges. For $R=1$~AU, the charge distribution is
practically symmetric with respect to the zero state. This symmetry is a characteristic feature of DD plasmas, where the
singly-charged positive and negative grains are the dominant charge carriers. For $R=100$~AU, the charge distribution is
slightly shifted to the negative values, representing DI plasmas. Low grain charges in the MRN case are due to a large
number of small ($a\lesssim0.01~\mu$m) grains -- they effectively reduce the abundance of free electrons and prevent larger
(sub-micron) grains from being multiply charged. However, as small grains are expected to disappear rapidly during the
initial stages of coagulations, larger grains become dominant in the size distribution.

For the evolved MRN distribution in Figure~\ref{fig4}, the average grain charge exhibits a linear scaling with the size; the
charge may be as high as $-50$ for $1~\mu$m grains. We notice that the results for $R=1$~AU and 100~AU represent,
respectively, DI plasmas and EI plasmas (as follows from the lower right panel of Figure~\ref{fig3}), so their direct
comparison may appear counterintuitive: for otherwise the same parameters, dust charges in DI plasmas should be lower than
in EI plasmas. The observed ``discrepancy'' is, however, due to the fact that temperatures are higher at smaller $R$.

The derived charge distributions show that one should be careful assuming low-charge states for protoplanetary disk
conditions. A moderate increase of the average grain size (above $\sim1~\mu$m) can easily break this assumption.

\subsection{Discussion}
\label{discussion}

The ``dead zone'' is a region of the protoplanetary disk where the development of MRI is suppressed by non-ideal MHD effects
\citep[e.g.,][]{Armitage2007}. It is well known that the size of the dead zone should strongly depend on dust properties
\citep{Sano2000,Bai2011a,Bai2011b,Dudorov2014}. The inner boundary of the dead zone is associated with the thermal
ionization near the central star, and is usually located at distances well below 1~AU \citep[][]{Chatterjee2014}. To roughly
estimate the position of the outer boundary, one can use the condition that the magnetic Reynolds number,
\begin{equation}\label{Reynolds}
 {\rm Re}_{\rm mag}=\frac{\alpha^{1/2} c_{\rm s}^2}{\eta_{\rm O}\Omega_{\rm K}},
\end{equation}
exceeds unity outside of the dead zone \citep[see, e.g., Equation~(170) of][]{Armitage2015}. Here, $\alpha\sim10^{-3}$ is
the turbulent alpha-parameter (outside of the dead zone), $c_{\rm s}=\sqrt{k_{\rm B}T/(2.3m_{\rm p})}$ is the sound speed,
and $\Omega_{\rm K}=\sqrt{GM_*/R^{3}}$ is the Keplerian angular velocity. For the estimates, we assume that the magnetic
diffusivity $\eta_{\rm O}$ at high gas densities is determined by the Ohmic term \citep[][]{Bai2016}, which can be
approximated by $\eta_{\rm O}=234(n_{\rm g}/n_{\rm e})\sqrt{T/{\rm K}}$ for a weakly ionized gas
\citep{SpitzerBook1962,Armitage2011}.\footnote{For the sake of simplicity, here we do not discuss the Hall and ambipolar
diffusion, as their role in the development of the dead zone requires more complex analysis \citep[e.g.,][]{Lesur2014}.}

To demonstrate the influence of the grain-size distribution on the size of the dead zone, in Figure~\ref{fig5} we plot the
abundances of the major charged species, calculated for different dust models. One can see that even a moderate increase in
the average grain size leads to significant changes in the ionization degree which, in turn, strongly affects the position
of the outer dead-zone boundary, marked by the vertical dashed line in each panel: The resulting size of the dead zone is
about 30, 10, 3 and 2~AU for, respectively, the MRN, monodisperse $0.1~\mu$m, evolved MRN, and monodisperse $10~\mu$m dust
size distribution.

As was pointed out in Section~\ref{window}, dust coagulation increases the ionization degree which, in turn, leads to higher
grain charges and prevents further coagulation due to growing electrostatic barrier. On the other hand, higher ionization
favors the development of MRI and thus stimulates the coagulation via turbulent dust motion. The situation become even more
complicated when the turbulence leads to the fragmentation of dust aggregates -- apart from destruction, this process
generates new populations of small dust which may reduce the electrostatic barrier and promote coagulation of larger grains.
Altogether, this suggests the existence of positive and negative feedback loops that may unpredictably halt or accelerate
the coagulation at different locations in the disk, and also highlights the importance of self-consistent analysis of the
ionization and dust evolution processes.

\section{Summary}
\label{conclusion}

We have developed an exact analytical model which describes ionization and dust charging in dense protoplanetary disk
conditions, for arbitrary grain-size distribution. Unlike previously developed approaches
\citep[][]{Ilgner2006,Okuzumi2009,Fujii2011,Dzyurkevich2013,Mori2016}, our model does not make assumptions on the form of
the grain charge distribution, and enables convenient analysis of results in a general form, in terms of a few dimensionless
numbers, which allows us to identify universality in the behavior of the charged species. The governing equations for
different cases are summarized in Appendix~\ref{notations}, Table~\ref{tabsum}.

For given dust properties and conditions of the disk, the presented model has only one free parameter (the effective mass of
the ions $A$), and is developed for the regime where the dust-phase recombination of free electrons and ions dominates over
the gas-phase recombination. A transition to this regime occurs in an electron-ion (EI) plasma (where charged grains still
do not play any role in the overall charge neutrality), and is characterized by the dust-phase recombination threshold
$(n_{\rm g}/\zeta)_{\rm rec}$ for the gas density.

At higher gas densities, $n_{\rm g}/\zeta\gtrsim(n_{\rm g}/\zeta)_{\rm rec}$, charged grains play an increasingly important
role in the charge neutrality. We have determined two characteristic parameters, the electron depletion threshold $(n_{\rm
g}/\zeta)_{\rm dep}\gg (n_{\rm g}/\zeta)_{\rm rec}$ and the asymptotic threshold $(n_{\rm g}/\zeta)_{\rm asy}\gg(n_{\rm
g}/\zeta)_{\rm dep}$, marking, respectively, transitions from the EI to dust-ion (DI) plasma state, and then to the
dust-dust (DD) state. The thresholds are determined in such a way that at $n_{\rm g}/\zeta=(n_{\rm g}/\zeta)_{\rm dep}$
electrons and negative grains equally contribute to the total negative charge, while at $n_{\rm g}/\zeta=(n_{\rm
g}/\zeta)_{\rm asy}$ ions and positive grains provide equal contribution to the total positive charge.

The immediate important implications of the derived results for protoplanetary disks are as follows:
\begin{enumerate}
\item Unless the dust size distribution is dominated by grains much smaller than $\sim1~\mu$m, larger grains are
    typically multiply (negatively) charged. In this case, irrespective of the location in the disk, the average grain
    charge scales linearly with the size. As the size distribution in protoplanetary disk conditions develops towards
    bigger grains, the conventional approximation of low grain charges may only be used for (very) initial stages of the
    disk evolution. The presented results are obtained assuming compact dust, and therefore the implication for porous
    aggregates (which are expected to carry higher charges due to bigger effective sizes) is even stronger.
    In situations where an aggregate is approximated by a sphere, the effects of porosity can be
    straightforwardly included by adopting a fractal scaling law $m_{\rm d}\propto a^D$, relating the dust mass and the
    effective size \citep[with appropriate fractal dimensionality $D<3$, which is known to vary during the dust
    evolution,][]{Okuzumi2009b}.
\item The asymptotic transition to a DD plasma implies that the grain charge distribution becomes quasi-symmetric with
    respect to $Z=0$. This completely removes the repulsive electrostatic barrier and opens a ``coagulation window'' for
    large aggregates, operating in the inner dense region of protoplanetary disks. On the other hand,
    the DD plasma state only ``delays'' the formation of the barrier: the coagulation itself leads to
    decreasing dust number density and a gradual shift back to DI/EI plasmas. The (re)appearance of the electrostatic
    barrier in this case, with the maximum achieved in the EI state (where the energy of the barrier normalized to the
    thermal energy of grains, $\sim |\langle Z\rangle|\Psi_{\rm EI}$, is always very large), may completely inhibit
    further dust growth. The effect of the barrier can be reduced by various feedback mechanisms operating in the disk
    and leading to increased local gas-to-dust ratio, such as the dust trapping or moderate fragmentation.
\end{enumerate}

We point out that the dust evolution, change in the charged species abundances, and development of MRI are strongly
interrelated processes whose mutual effect is poorly understood: The dust coagulation increases the ionization degree which,
in turn, leads to higher grain charges and prevents further coagulation due to growing electrostatic barrier. On the other
hand, higher ionization favors the development of MRI, making a disk turbulent; a moderate turbulence facilitates dust
growth by increasing the relative grain velocities, while strong turbulence leads to dust fragmentation. The latter
generates small grains, which may decrease the electron fraction (asymptotically, by a factor of $\sqrt{m_{\rm i}/m_{\rm
e}}\sim10^2$) and lead to MRI quenching. A complex interplay of these nonlinear processes suggests the existence of multiple
positive and negative feedback loops that may dramatically affect the ultimate dust evolution.

A rigorous treatment of the ionization fraction and dust evolution could be critical during all stages in the process that
links molecular clouds to stellar systems -- this is the motivation behind our work. The exact analytical model presented
here can be easily implemented in non-ideal MHD simulations, to properly follow the ionization fraction and the dust growth
during the process of protoplanetary disk formation \citep[e.g.,][]{Zhao2016} and evolution \citep[e.g.,][and references
therein]{Dullemond2007,Armitage2011}.

We provide a FORTRAN source code, applicable for arbitrary dust size distributions and also for porous grains, to calculate
abundances of the charged species: http://www.inasan.ru/\~{}akimkin/codes.html

\section*{Acknowledgements}

This research made use of NASA's Astrophysics Data System. The authors thank an anonymous referee for detailed comments and
suggestions, and Dr. Bo Zhao and Dr. Sergey Khaibrakhmanov for useful discussions. VVA acknowledges financial support from
Russian Foundation for Basic Research (16-32-00012 mol\_a) and from a grant of the President of the Russian Federation
(NSh-9576.2016.2). PC acknowledges financial support from the European Research Council (ERC Advanced Grant PALs 320620).

\appendix

\section{Appendix A: Summary of governing equations}
\label{notations}

For convenience, in Table~\ref{tabsum} we list the equations to be used in a general case, as well as in different limiting
cases.

\begin{table*}[!ht]\centering
\caption{}\label{tabsum}
   \begin{tabular}{| l | l | l | l | l | }
     \hline
      Dust size distribution; limiting case
      & Parameters
      & Governing equations
      & Auxiliary relations
      & Solution for
      \\ \hline
      Polydisperse; arbitrary $\tf$
      & $n_{\rm g}, \zeta, f_{\rm d}, T, A, dn_{\rm d}/da$
      & \eqref{ion-rec}--\eqref{norm1}, \eqref{distr-}, \eqref{distr}
      & \eqref{fraction}, \eqref{norm}, \eqref{parameter1}, \eqref{parameter2}
      & $n_{\rm i}, n_{\rm e}, N_{Z}$
      \\ \hline
      Monodisperse; $\tf\ll1$
      & $n_{\rm g}, \zeta, f_{\rm d}, T, A, a$
      & \eqref{Psi}, \eqref{gov1a}, \eqref{gov1b}
      & \eqref{fraction}, \eqref{parameter1}, \eqref{parameter2}, \eqref{Gauss}
      & $n_{\rm i}, n_{\rm e}, \langle Z\rangle$
      \\ \hline
      Monodisperse; $\tf\gg1$
      & $n_{\rm g}, \zeta, f_{\rm d}, T, A, a$
      & \eqref{N0}--\eqref{gov2a}, \eqref{gov2b}
      & \eqref{fraction}, \eqref{parameter1}, \eqref{parameter2}, \eqref{discrete}, \eqref{2_mch}
      & $n_{\rm i}, n_{\rm e}, N_0, N_{-1}$
      \\ \hline
   \end{tabular}
\end{table*}

\section{Appendix B: Grain charge distribution $N_{Z}$}
\label{charge_dist}

When the photoemission from grains as well as other emission mechanisms are negligible, $N_{Z}$ is determined by the
collection of electrons and ions from the ambient plasma. The sticking probabilities of electrons and ions are both assumed
equal to unity. The collection cross sections, determined by the electrostatic interactions with a charged grain, are
derived from the OML approximation \citep[][]{Whipple1981,Fortov2005}. For positive charge states, the detailed equilibrium
yields \citep[][]{Draine1987,Draine2011Book}
\begin{equation}
Z\geq0:\quad\frac{N_{Z+1}}{N_{Z}}=\left(\frac{e^{-Z\tf}}{1+(Z+1)\tf}\right)\frac1{\sqrt{\tm}},\label{distr+}
\end{equation}
negative charge states are related to the respective positive states via
\begin{equation}
N_{-Z}=\tm^{Z}N_{Z}.\label{distr-}
\end{equation}
From Equation~(\ref{distr+}) we derive
\begin{equation}\label{distr}
Z>0:\quad\frac{N_{Z}}{N_0}=\frac{e^{-\frac12Z(Z-1)\tf}}{\tm^{Z/2}\prod_{Z'=1}^Z(1+Z'\tf)},
\end{equation}
$N_{-Z}$ is readily obtained by using Equation~(\ref{distr-}). Note that the charge distribution is slightly modified when
the polarization interactions are taken into account \citep[e.g., Equations~(3.3) and (3.4) in][]{Draine1987}. If needed,
this effect can easily be included in Equations~(\ref{distr+})--(\ref{distr}) \citep[][]{Ivlev2015b}. To include the
stickling probabilities of electrons, $s_{\rm e}(Z)$, and ions, $s_{\rm i}(Z)$, the rhs of Equation~(\ref{distr+}) should be
multiplied by the ratio $s_{\rm i}(Z)/s_{\rm e}(Z+1)$.

\section{Appendix C: Effective dust density $\mathcal{N}$ for the ion collection}
\label{app_flux}

The effective number density $\mathcal{N}(a)$ of grains of radius $a$, entering the dust-phase recombination term $R_{\rm
d}$ in Equation~(\ref{R_id}), takes into account the electrostatic interaction between ions by charged grains. Depending on
the sign of the grain charge, the (geometrical) cross section of the ion collection by the grain is increased or decreased;
the corresponding factors are derived from the OML approximation. Summing up partial contributions of all charged states
yields
\begin{equation*}
\mathcal{N}(a)=\sum_{Z<0}(1-Z\tf)N_{Z}+\sum_{Z\geq0}e^{-Z\tf}N_{Z}.
\end{equation*}
We use Equation~(\ref{distr-}) to eliminate the summation over positive charges, and employ the recurrent relation
(\ref{distr+}) to rewrite $\mathcal{N}$ in the identical form,
\begin{equation}\label{Neff}
\mathcal{N}(a)=\sum_{Z\geq0}(\sqrt{\tm}+\tm^{-Z})e^{-Z\tf}N_{-Z},
\end{equation}
which is more convenient for the analysis (see Sections~\ref{tf<<1} and \ref{tf>>1}).

\bibliographystyle{apj}
\bibliography{refs}

\end{document}